\documentclass{aa}
\usepackage{graphicx}
\begin{document}


   \title{The red tail of carbon stars in the LMC: \\ 
Models meet 2MASS and DENIS observations}
 
   \author{P. Marigo$^1$, L. Girardi$^2$, \& C. Chiosi$^1$}

   \institute{$^1$ Dipartimento di Astronomia, Universit\`a di Padova, 
		Vicolo dell'Osservatorio 2, 35122 Padova, Italy \\
              $^2$ Osservatorio Astronomico di Trieste, Via Tiepolo 11, 
                34131 Trieste, Italy}

   \titlerunning{The red tail of carbon stars in the LMC}
   
   \offprints{P. Marigo \\ \email{marigo@pd.astro.it}}

   \date{To appear in A\&A, see http://pleiadi.pd.astro.it for a version with 
high-resolution figures.}

   \abstract{
Carbon stars are known to exhibit systematically redder near-infrared
colours with respect to M-type stars.  In the near-infrared
colour--magnitude diagrams provided by the 2MASS and DENIS surveys,
the LMC C-type stars draw a striking ``red tail'', well separated from
the sequences of O-rich giants. So far, this conspicuous feature has
been absent from any set of available isochrones, even the few
existing ones that include the TP-AGB evolution of low- and
intermediate-mass stars. 
To investigate such issue we simulate  the complete 
2MASS $K_{\rm s}$ vs. $(J-K_{\rm s})$ data towards the LMC 
by means of a population synthesis approach, that relies on extended
libraries of published stellar evolutionary tracks, including
the TP-AGB phase.
The simulations provide  quite a detailed description of the several
vertical ``fingers'' and inclined sequences seen in 2MASS
data, due to both Galactic foreground and LMC O-rich stars. Instead,
as mentioned, the red tail of C-stars sets a major difficulty: we find
that TP-AGB models with solar-scaled molecular opacities, the usual
assumption of existing AGB calculations, do not succeed in reproducing
this feature.
Our tests indicate that the main reason for this failure should 
not be ascribed to empirical $T_{\rm eff}$--$(J-K)$ transformations 
for C-type stars.
Instead, the discrepancy is simply removed
by adopting new evolutionary models that account for the changes 
in molecular opacities as AGB stars get 
enriched in carbon  via the third dredge-up (Marigo 2002). 
In fact, simulations that adopt these models are able to reproduce,
for the first time, the red tail of C-stars in near-infrared CMDs. 
Finally, we point out that these simulations also provide 
useful indications about the efficiency of the third dredge-up
process, and the pulsation
modes of long-period variables.
   }
	
   \maketitle
%

\section{Introduction}
It has been already known for a long time that carbon 
(C-type) stars occupy a particular region of near-infrared
colour-colour diagrams (e.g. Richer et al. 1979; Cohen et al. 1981; 
Frogel et al. 1990; Hughes \& Wood 1990). 
Specifically, C-type stars exhibit systematically
redder colours than M-type stars, and particular 
colour-colour diagrams (e.g. $J-H$ vs.\ $H-K$) 
display such a sharp dichotomy between 
the two spectral classes that they are often used diagnostic 
tools to identify carbon star candidates from observed samples. 

\begin{figure*}
\begin{minipage}{0.48\textwidth}
\includegraphics[width=\textwidth]{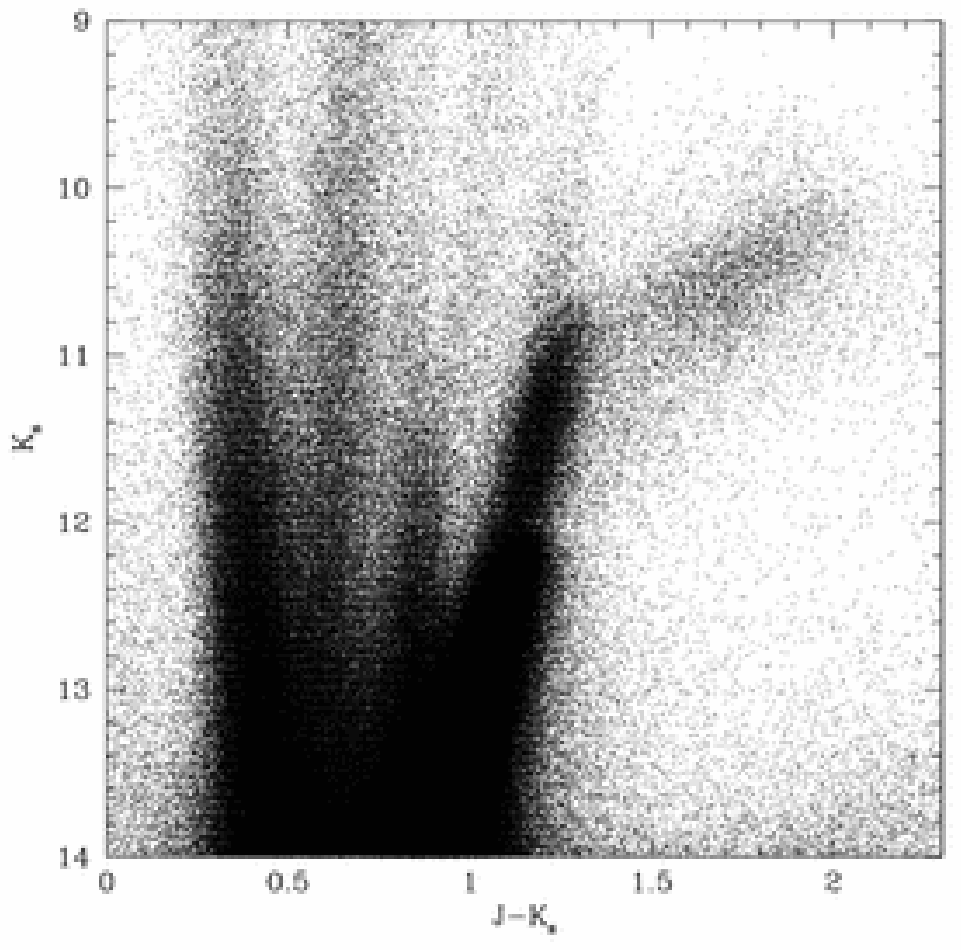}
\end{minipage}
\hfill
\begin{minipage}{0.48\textwidth}
\includegraphics[width=\textwidth]{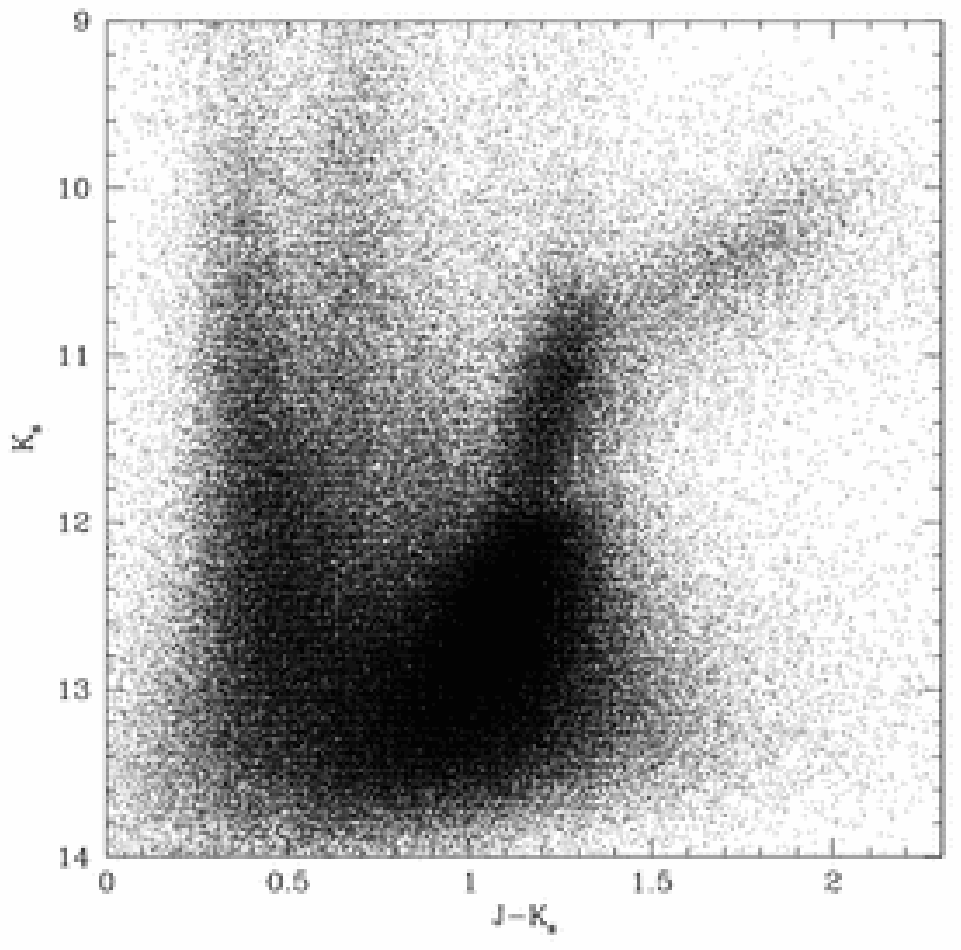}
\end{minipage}
\caption{ {\bf Left panel:} $K$ vs. $(J-K)$ diagram for the 
LMC region of sky ($101^\circ\le\alpha\le61^\circ$ and 
$-77^\circ\le\delta\le-63^\circ$). The data are taken from 
the 2MASS Second Incremental Data Release.
Note the red tail of visible C-stars departing from 
the sequence of red giants at $(J-K)\sim1.4$ 
and extending up to $(J-K)\sim2$. 
{\bf Right panel:} The same from DENIS Point Source Catalogue
towards the LMC (Cioni et al. 2000).}
\label{fig_denis2mass}
\end{figure*}

Recent wide-area near-infrared surveys like 
the Two Micron All Sky Survey (2MASS; Skrutskie et al. 1997) 
and the Deep Near-Infrared Southern Sky Survey 
(DENIS; Epchstein et al. 1997) have confirmed these
findings. By providing photometric data for 
the complete AGB population of the Magellanic Clouds,
they reveal  the presence  of a red plume of C-stars in a striking way 
(Cioni et al. 1999; Nikolaev \& Weinberg 2000).
This feature can be appreciated in the $K_{\rm s}$ vs.\ $(J-K_{\rm s})$ 
diagrams of Fig.~\ref{fig_denis2mass}: they show
a marked almost-vertical sequence of red giants at $(J-K_{\rm s})\sim1.1$, 
from which an inclined 
branch departs at $K_{\rm s}\sim11$ towards redder colours,
extending up to $(J-K_{\rm s})\sim2.0$.
This inclined branch is what we refer to as ``the red tail''.
Nikolaev \& Weinberg (2000) and Cioni et al. (2001) demonstrate 
that the red tail is populated by non-obscured
C stars, whereas the almost-vertical
feature at $(J-K_{\rm s})\sim1.1$ is composed basically of 
O-rich giants (of spectral types from late-K to M). 
Moreover, the luminosities of red tail stars indicate that they 
belong to the thermally pulsing asymptotic giant branch (TP-AGB) 
phase, and hence they are expected to be the result of carbon surface 
enrichment by the recurrent third dredge-up process. 

As for the modelling of this feature, the present-day situation is 
as follows. Most of the stellar isochrones
for old and intermediate ages available in the literature  
do not include the TP-AGB phase, as they usually extend up to 
tip of the red giant branch (RGB) or the Early-AGB
(e.g. Lejeune \& Schaerer 2001; 
Dom\'\i nguez et al. 1999; Weiss \& Salaris 1999; 
Bergbusch \& VandenBerg 2001 and references therein).
Two main reasons for this discrepancy 
are that i) fully modelling the AGB phase is
indeed complex and time-consuming, and ii) most stellar models
fail to explain the formation of carbon stars {except}
for those at higher luminosities.   

\begin{figure}
\includegraphics[width=\columnwidth]{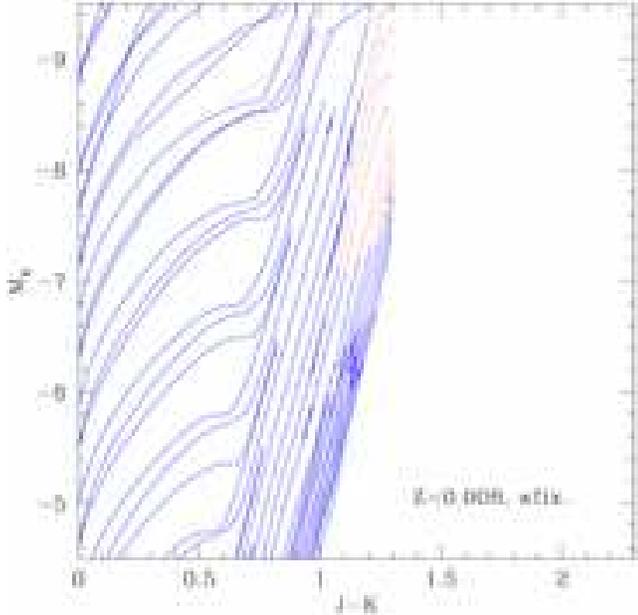}
\caption{A subset of theoretical isochrones described in
Marigo \& Girardi (2001), and transformed to $JK$ photometry
by means of Girardi et al. (2002) tables. 
TP-AGB models are computed with the $\kappa_{\rm fix}$ prescription.
The metallicity is $Z=0.008$, and isochrones are presented at
equally-spaced age intervals of $\Delta \log t = 0.1$.
The dotted lines 
mark the isochrone sections corresponding to C-stars (with C/O$>1$). 
The plot limits correpond closely to those of 
Fig.~\protect\ref{fig_denis2mass} for a LMC distance modulus
of 18.5~mag. In these and other models in literature, 
C stars do not form any red tail.
}
\label{fig_isockfix}
\end{figure}

To our knowledge the only available isochrone sets that include TP-AGB stars 
are the Padova ones (Bertelli et a. 1994; Girardi et al. 2000; 
Salasnich et al. 2001;  Marigo \& Girardi 2001). 
In these models the entire TP-AGB evolution is followed -- 
 with the aid of a synthetic approach to various degrees of detail -- 
up to ejection of the stellar envelope. 
However, all of them  fail to predict the 
red tail of C stars, and in fact the reddest TP-AGB
tracks hardly reach $(J-K)$ colours as large as 1.3. This is 
illustrated in Fig.~\ref{fig_isockfix}. 
If in the older isochrone sets (Bertelli et al. 1994; Girardi et al. 2000) 
the basic problem was the lack of the C-star phase in 
the stellar models, the same does not affect the most recent 
models (Marigo \& Girardi 2001), that account for the formation of C-stars 
via the third dredge-up.

What is then the problem with the colours of
theoretical C-star models?
Why do they not reach $(J-K)$ as red as $\sim2$\,? 
Since in Marigo \& Girardi (2001) isochrones 
the third dredge-up process is calibrated
so that C-stars appear at the right luminosities (see Marigo et
al. 1999 for details), the problem might be related either to
(i) an improper transformation from $T_{\rm eff}$ to the $(J-K)$
colour, which introduces errors of several tenths of a magnitude
for C-type stars, and/or 
(ii) an improper modelling of the C-star radii and effective 
temperatures, with typical errors of some hundreds of degrees
Kelvin. 

In case of alternative (i), a simple revision of the 
$T_{\rm eff}$--colour  relations, via e.g. the
use of suitable relations for M- and C-type stars, would 
solve the problem.
In the case of option (ii), the implications would be much deeper, 
since the $T_{\rm eff}$ values of Marigo et al. (1999)
models are typical of most TP-AGB models in the literature.
 
Exploring both possibilities is the main scope of this paper, which
is organised as follows.
In Sect.~\ref{sec_Cstars} we start {by} recalling the main prescriptions
in our synthetic TP-AGB models, that also account for the transition
from O-rich (M-type) to C-rich (C-type) stars (Sect.~\ref{ssec_modpre}).
A brief summary {of} the present status of other AGB evolution models 
available in the literature is given in Sect.~\ref{ssec_sum}, putting 
particular emphasis on the role of molecular opacities.

Section~\ref{sec_comparison} describes our simulations of  
the 2MASS data in the $K$ vs. $(J-K)$ colour magnitude diagram (CMD). 
If  several 2MASS features are reproduced in detail, 
the red tail of C-stars is missed by our initial models.
Section~\ref{sec_probing} investigates the possible reasons for it.
Finally, our main conclusions are summarised in Sect.~\ref{sec_conclusion}.

\section{Models for C-stars}
\label{sec_Cstars}
\subsection{Our prescriptions: old and new}
\label{ssec_modpre}
In this study the formation and evolution of C-stars is described 
with the aid the {\sl synthetic} TP-AGB model developed by 
Marigo et al. (1996, 1998, 1999), and Marigo (2002) to whom the reader
is referred for all the details.
For the sake of clarity let us  briefly outline the main structure 
of the model and its basic prescriptions.

\begin{itemize}
\item The TP-AGB evolution -- of a star of given initial
mass and metallicity -- is followed from the first thermal pulse
up to the complete ejection of the envelope by stellar winds.
This is performed by suitably coupling (i) analytic prescriptions
(e.g. the core mass-luminosity relation,  the core mass-interpulse relation,
the growth rate of the core mass during the quiescent stages), and
(ii) a complete static envelope model to be integrated from the atmosphere
down to the core.
\item The initial conditions at the first thermal pulse are derived from
sets of stellar tracks for low- and intermediate-mass stars 
by Girardi et al. (2000), which
cover the previous evolutionary phases, i.e. from the zero-age main sequence
up to the beginning of the TP-AGB phase.
In this way consistency and homogeneity are guaranteed in the construction of
the corresponding stellar isochrones (see Sect.~\ref{sec_simulating}).
\item Third dredge-up episodes possibly take place at thermal pulses 
and eventually lead to the transition from M- to C-class if 
the surface C/O ratio increases above 1. 
The third dredge-up is described as a function of two 
parameters, namely: the efficiency $\lambda$, and the minimum temperature
$T_{\rm b}^{\rm dred}$ at the base of the convective envelope
for dredge-up to take place.
The previous calibration of these quantities 
(Marigo et al. 1999) on the observed 
luminosity functions of C-stars has yielded ($\lambda=0.50, \, 
\log T_{\rm b}^{\rm dred} = 6.4$) for the LMC and  ($\lambda=0.65, \, 
\log T_{\rm b}^{\rm dred} = 6.4$) for the SMC.
\item Hot-bottom burning occurring in the most massive AGB stars
($M \ga 4.5\, M_{\odot}$), is followed in detail with the 
aid of complete envelope integrations (Marigo et al. 1998, Marigo 1998).
The possible break-down of the core mass-luminosity relation 
and the CNO-cycle nucleosynthesis (possibly preventing the 
formation of luminous C-stars) are both taken into account.
\item Mass loss is included according to Vassiliadis \& Wood's (1993) 
formalism, based on observations of pulsating AGB stars.
In our calculations we assume that pulsation occurs with 
periods ($P$) corresponding to either the fundamental mode ($P_0$), 
or the first overtone mode ($P_1$). 
\item Of great importance for the analysis developed in this work are
the opacity prescriptions to be used in the envelope model. 
At high temperatures
($T\ge10000$ K) we adopt Iglesias \& Rogers (1993)
for O- and C-rich mixtures. 
At low temperatures ($T<10\,000$ K) we use either the solar-scaled
opacity tables  by Alexander \& Ferguson (1994; hereafter also 
$\kappa_{\rm fix}$), or the routine developed by 
Marigo (2002; hereafter also $\kappa_{\rm var}$) that predicts 
molecular opacities for any chemical composition of the gas.
It should be remarked that before Marigo (2002), all TP-AGB tracks
by Marigo et al. were calculated with $\kappa_{\rm fix}$. 
\end{itemize}

In summary, for the purpose of our present work  we will make use of two
groups of synthetic TP-AGB calculations, namely:
\begin{itemize}
\item Old TP-AGB models with $\kappa_{\rm fix}$ (Marigo et al. 1999); 
\item New TP-AGB models with $\kappa_{\rm var}$ (Marigo 2002).
\end{itemize}
For the remaining prescriptions, we refer to the aforementioned
model outline unless otherwise specified.  

\subsection{The formation of C-stars in other models 
and the role of molecular opacities}
\label{ssec_sum}
Before proceeding with our analysis it is worth recalling briefly
the current status of  AGB models available in
the literature and the importance of a {consistent} choice in 
the use of molecular opacities.

First of all, the lack of efficient dredge-up 
in low-mass AGB models (say with $M \la 2 M_{\odot}$) has been a 
difficulty shared by all full stellar evolution codes for a long time
(the so-called ``carbon star mystery'', as referred to by Iben (1981)).
More recently this difficulty seems to be overcome by {a} few groups
(e.g. Herwig et al. 1997; Straniero et al. 1997; Karakas et al. 2002),
due to different numerical treatments in describing 
the convective boundaries.

However, {even} if some progress is achieved by full stellar evolution
models in relation to the formation of low-mass C-stars, the distance
between theory and observations is still large.
{Currently} there is no complete set of evolutionary
tracks -- based on full stellar models --  extending up to the end
of the AGB phase for a sufficient coverage of stellar masses and 
metallicities. This prevents a systematic test of stellar models
by comparison with observations, e.g. the luminosity functions of 
M- and C-stars, and simulations of CMDs including the most
evolved AGB stars.
The main {reason}  is the heavy
computational effort required {to  fully model} the AGB phase.
On the other hand, this limit can be totally removed if one 
allows for a certain loss of {detail} and opts for the more agile 
synthetic approach (see e.g. Groenewegen \& de Jong
1993, Marigo et al. 1996, 1998 for recent works).

Another heavy inadequacy of present stellar models 
(both full and synthetic) {that} has been 
recently highlighted by Marigo (2002) is the use of 
fixed solar-scaled molecular opacities 
in AGB models experiencing the third dredge-up.

\begin{figure}
\includegraphics[width=\columnwidth]{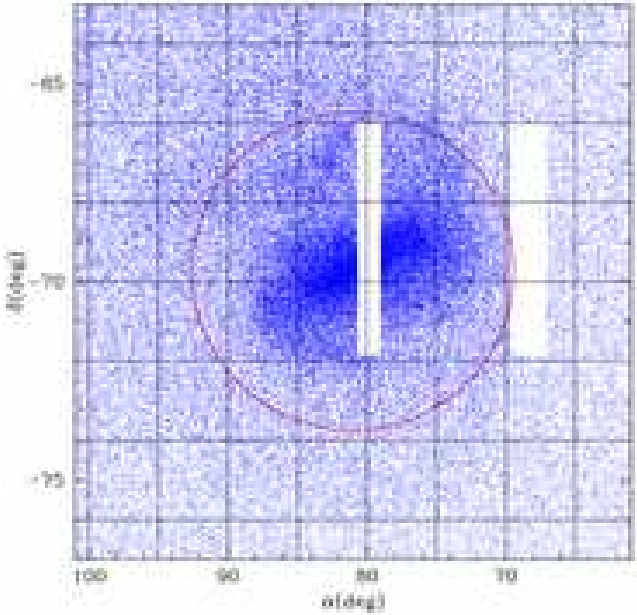}
\caption{Distribution of point sources from the
2MASS Second Incremental Data Release in the LMC region of 
the sky, projected on the $(\alpha,\delta)$ plane. 
To reduce confusion, we plot only the objects with $K_{\rm s}<12.5$.
The circle delimits the area selected for our analysis. }
\label{fig_map}
\end{figure}

\begin{figure}
\includegraphics[width=\columnwidth]{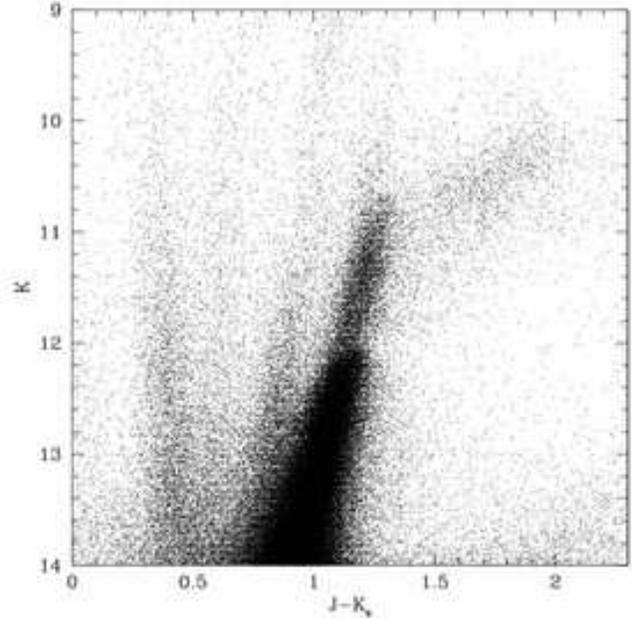}
\caption{2MASS data for the innermost $4^\circ$ 
of the LMC. Note the red tail of
visible C-stars extending up to $(J-K_{\rm s})\sim 2$. }
\label{fig_cmdfinal}
\end{figure}

In fact, this prescription 
(previously defined as $\kappa_{\rm fix}$ case) 
is still the standard choice common to most 
{\sl full and synthetic} evolutionary
calculations\footnote{Actually, non-solar scaled molecular opacities 
were already  taken into account in a simple way by Vassiliadis \&
Wood (1993), but no significant difference emerged in comparison to models
with solar-scaled opacities, as the authors  did not practically find
carbon dredge-up in their AGB calculations.}
of the AGB phase (e.g. Mouhcine \& Lan\c con 2002; 
Chieffi et al. 2001; Herwig 2000; 
Ventura et al. 1999;
Wagenhuber \& Groenewegen 1998;
Forestini \& Charbonnel 1997; Straniero et al. 1997).

Marigo (2002) has proved, instead,  that properly coupling 
the molecular opacities to the actual surface chemical abundances
($\kappa_{\rm var}$) brings along such important 
consequences that the  standard evolutionary scenario for the AGB evolution 
may be significantly affected. The reader is referred to that work 
for a detailed analysis.   
  
One major effect is that the surface enrichment in carbon 
that leads to C/O$>1$ -- i.e. the formation of carbon stars -- 
causes a significant increase in opacities, which in turn 
is responsible for a sudden and marked cooling of the stellar 
tracks in the H-R diagram.
The sizeable excursion towards lower $T_{\rm eff}$, displayed by
the $\kappa_{\rm var}$  C-star tracks, 
sets the first significant difference compared to $\kappa_{\rm fix}$ 
C-star models that, instead, do not show any change in the slope of
their evolutionary tracks while climbing the AGB at increasing  luminosities. 

A closer comparison between models with $\kappa_{\rm fix}$ and 
$\kappa_{\rm var}$, and the impact of the new opacities prescription
will be discussed in Sect.~\ref{sec_testtracks}.

\section{The $K$ vs. $(J-K)$ diagram towards the LMC}
\label{sec_comparison}

In this section, we aim to model the red tail of carbon stars
in the LMC by using the TP-AGB models at our disposal. 
Our study is limited to a particular CMD -- the $K$ vs. $(J-K)$ one
-- that reveals the red tail in all its prominence
(see Fig. \ref{fig_denis2mass}). Needless to say, the same 
feature is displayed in a variety of different CMDs involving 
near-infrared passbands, as can be seen e.g. in  
van der Marel \& Cioni (2001, their figure 2).

\subsection{The selected data}
\label{sec_data}
In the following, we base our comparisons on 2MASS data towards
the LMC, that present an extremely good photometric quality.
More specifically, we use the point-source catalog of the 
2MASS Second Incremental Data Release (see Cutri et al. 2002), 
which includes $10\sigma$ detections as faint as 
$16.3$, $15.3$, $14.7$ mag in $J$, $H$ and $K_{\rm s}$, respectively. 
As it can be seen in Fig.~\ref{fig_map}, 
these data wholly cover the LMC except for two limited strips of 
right ascension. In order to reduce the fore-/back-ground 
of Galactic stars and distant galaxies,
we select only the stars located closer
than 4 degrees (projected) to the LMC bar optical center 
(at $\alpha=5^{\rm h}24^{\rm m}$, $\delta=-69^{\circ}44\arcmin$, 
cf.\ Nikolaev \& Weinberg 2000). 
This region, also considering the empty strip, 
translates into to a total sky area of about 48.2 squared degrees.
The corresponding CMD is shown in Fig.~\ref{fig_cmdfinal}.

\subsection{Simulations}
\label{sec_simulating}
The $K$ vs. $(J-K)$ diagram has been simulated by using a
population synthesis code (Girardi et al., in preparation).
In short, the code randomly generates stars
following a given star formation rate (SFR), 
age-metallicity relation (AMR) and initial mass function 
(IMF). The stellar intrinsic properties (luminosity $L$,
effective temperature $T_{\rm eff}$, surface gravity $g$, 
etc.) are interpolated
over a large grid of stellar evolutionary tracks, based on
Bertelli et al. (1994) for massive stars,
Girardi et al. (2000) for low- and intermediate-mass stars,
and complemented with the TP-AGB tracks from either
Marigo et al. (1999) with $\kappa_{\rm fix}$, 
or new calculations with $\kappa_{\rm var}$ 
made on purpose (refer to Marigo 2002; see
also Sect. \ref{sec_Cstars}). 

The evolution along the TP-AGB phase is initially described in
terms of the properties in the quiescent stages between thermal
pulses. Actually, there are two additional factors that 
cause significant excursions in luminosity and effective temperature
from their quiescent values, namely:
\begin{itemize}
\item occurrence of thermal pulses;
\item instability against pulsation.
\end{itemize}

The former effect is included in our simulations. 
In practice, the luminosity variations driven by a thermal pulse
(both the fast luminosity peak and long-lasting 
low-luminosity dip) are described according to the luminosity
distributions over a pulse cycle as derived from 
Boothroyd \& Sackmann (1988), and 
assuming that the envelope mass is the main factor determining 
the shape of the luminosity dip. Once a star is randomly 
scattered in $\log L$ according to this distribution, 
the corresponding excursion in effective temperature is 
determined from a grid of envelope integrations performed in 
precedence. In this way, the intrinsic dispersion of TP-AGB stars
in the HR diagram is realistically simulated with just a 
modest computational cost. 
We recall that this dispersion is important mainly for 
low-mass AGB stars, with a maximum displacement of
$\approx -0.4$ in $\log L$ (or $\sim +1$ mag), affecting up to 30 percent 
of their TP-AGB lifetimes. 

As for the latter point related to stellar pulsation, 
it is worth premising the following.
We recall that both 2MASS and DENIS 
provide single-epoch observations. 
Hence, the sampling of variable stars 
(e.g. Miras, SR variables) at a single and 
random phase of their variability cycle causes a further dispersion of 
data points in the observed CMD.
This effect has not been 
simulated here for two reasons. First, we do not 
know exactly how $L$ and $T_{\rm eff}$ vary
during the pulsation cycle. Second, in any case the
effect of this cyclical variation is easy
to foresee: the bulk of TP-AGB stars will be 
scattered in the CMDs, to both higher and lower 
magnitudes and colours. 
The expected result is a sort of general
blurring of the TP-AGB sequences in the CMD.
Mira and SR variables have pulsation amplitudes ranging
from 0.4 to 1.0 mag in the $I$-band (Hughes \& Wood 1990).
Therefore, to take this effect into account the 
simulated TP-AGB sequences should be blurred by 
just some tenths of magnitude. 

Once the stellar intrinsic properties are singled out,
the photometry is generated by applying the extended tables
of bolometric corrections from Girardi et al. (2002) for
O-rich stars (with C/O$<1$), and empirical relations -- to be discussed
in Sect.~\ref{sec_testcolours} below -- for C-type stars (with C/O$>1$). 
\begin{figure*}
\begin{minipage}{0.48\textwidth}
\includegraphics[width=\textwidth]{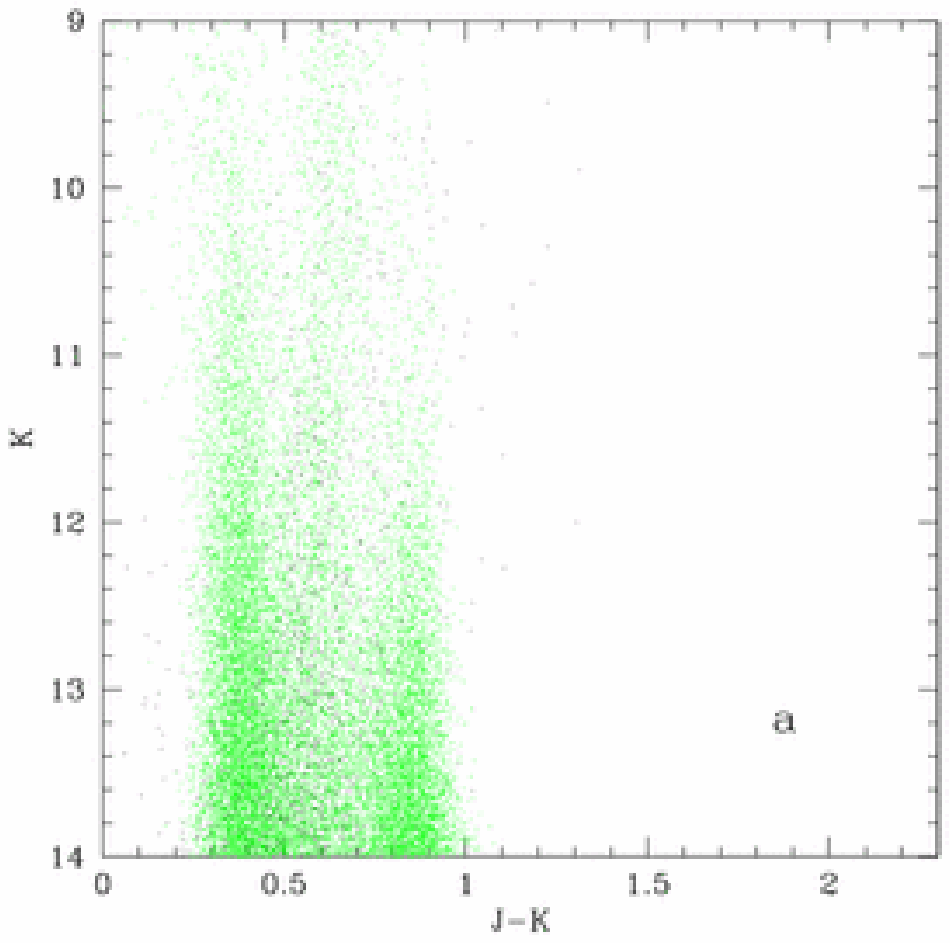}
\end{minipage}
\hfill
\begin{minipage}{0.48\textwidth}
\includegraphics[width=\textwidth]{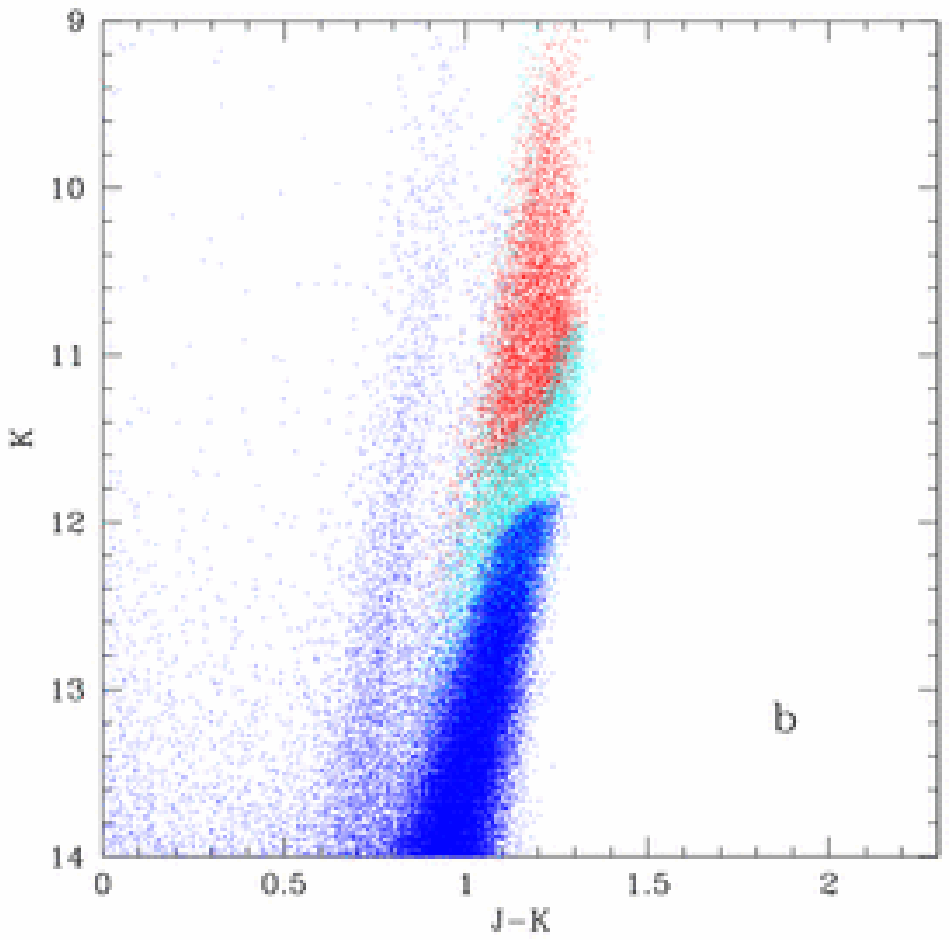}
\end{minipage}
\caption{ 
Simulated CMDs for both the Galaxy foreground and the LMC population. 
{\bf Panel a:} The Galaxy foreground stars. The coloured dots mark 
the stars belonging to the disk (green) and halo (black) populations.
{\bf Panel b:} The LMC population, marked in blue
for all stars before the TP-AGB phase, and separated into O-rich 
(cyan) and C-rich stars (red) during the TP-AGB phase. 
Note that the C-star population distribute almost vertically over 
the red giant branch. 
}
\label{fig_comp0}
\end{figure*}

Our simulations are computed in the $JHK$ filter system defined by 
Bessell \& Brett (1988).
The expected differences from the 2MASS and DENIS systems --
that use a ``$K$-short'' filter, 
$K_{\rm s}$ -- are probably very low, of less than say 
0.1 mag in $(J-K)$. This is illustrated in the case of 2MASS by 
Cutri et al. (2002, section II,2d). 

We simulate the photometric errors by means of a Monte
Carlo approach, adopting  the distribution of error values tabulated 
from 2MASS.  We recall that $\sigma$ errors are lower than 
$0.03$~mag for $K_{\rm s}\la13$.

\begin{figure}
\includegraphics[width=\columnwidth]{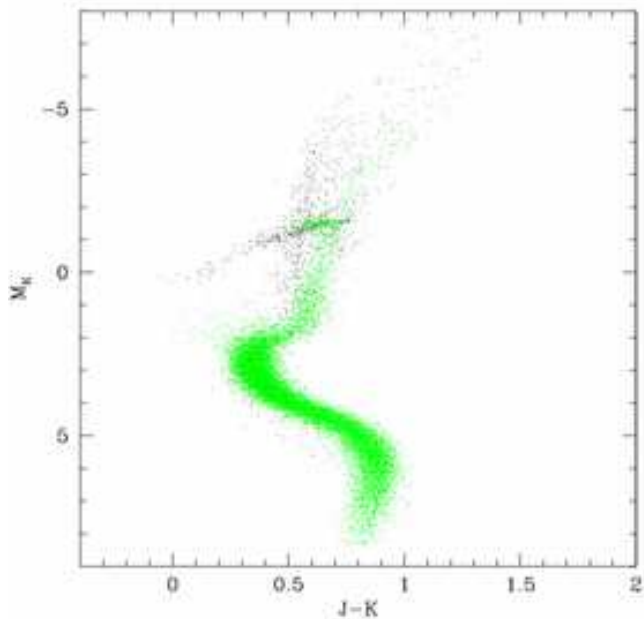}
\caption{ 
The same data as in Fig. \protect\ref{fig_comp0}a,
but in the $M_{K}$ vs. $(J-K)$ plane.
}
\label{fig_comp0exp}
\end{figure}

Finally,
the population synthesis code generates simulations for 
both the LMC galaxy  and the Galactic {foreground}, 
by using different distributions of reddening and distances.
Results are shown in Fig.~\ref{fig_comp0},
for the specific case of TP-AGB models computed with 
solar-scaled molecular opacities ($\kappa_{\rm fix}$), and 
discussed in the following.

\subsubsection{The Galactic foreground}

The field foreground stars have been included with the aid of
a complete Galaxy model (Girardi et al., in preparation; 
Groenewegen et al. 2002). 
It consists of disk and halo components:

The disc is described by
a double-exponential density law, with a radial scale length (in 
Galactocentric distance) of 2.8 kpc, and a scale height $H$ that 
increases with stellar age $t$ as
\[
H = z_0 \, (1+t/t_0)^\alpha \,\,\,.
\]
We adopt the parameters $z_0=95$ pc, $t_0=0.5$ Gyr and $\alpha=1.66$, 
that well describe the disk components as derived by Ng et al. (1997). 
It is worth noticing that this prescription allows for a significant 
increase of $H$ at large ages, which is similar to assuming the 
presence of an old thick disk component. 
The distributions in mass, age and metallicity of the simulated
disk stars are derived under the assumptions of:
a constant star formation rate {between 
12 Gyr ago and now}; the empirical age-metallicity relation from 
Rocha-Pinto et al. (2000) with the addition of a dispersion
of 0.2 dex in $[{\rm Fe/H}]$ at any age; and the log-normal IMF from 
Chabrier (2001).

The halo is described by an oblate spheroidal 
(cf. Gilmore 1984) of axial ratio $q=0.65$, and a core radius
of 2.8 Kpc. Halo stars are assumed to be old (from 12 to 
13 Gyr), and to follow a Gaussian distribution of metallicities
of mean $[{\rm Fe/H}]=-1.6$ and standard deviation of 1.0 dex. 

This set of prescriptions has been constrained by 
Groenewegen et al.\ (2002), by
using a set of Deep Multicolor Survey (Osmer et al.\ 1998)
and ESO Imaging Survey data (Prandoni et al.\ 1999).
Interestingly, most of the calibration data
were in visible pass-bands, whereas the model  is now being 
applied to the interpretation of near-infrared data.

For the purpose of our study,
we simulate a conic section of the Galaxy occupying  
the same sky area as in the selected 2MASS data.
The field center has galactic coordinates 
$\ell=280^{\circ}.46$, $b=-32^{\circ}.89$. We recall that, for
this particular line-of-sight, other Galactic components 
(spiral arms, the bulge) are of no relevance.

An important aspect is that in the modelling we can distinguish  
different kinds of stars and hence tackle  the origin of
the several observed features in the CMD of 2MASS.

Our field simulation is presented in
(Fig.~\ref{fig_comp0}a). In this plot, 
we can identify three marked vertical sequences 
of foreground disk stars (green dots), located at $(J-K)$
colours of about 0.35, 0.65, and 0.9. Their origin is easily
understood by looking at the corresponding plot in absolute
magnitude (Fig.~\ref{fig_comp0exp}): 
The sequence to the left (at $(J-K)\simeq0.35$) is defined 
mainly by the old disk turn-off, with typical masses 
of $M\simeq0.9$~$M_\odot$; the central one (at $(J-K)\simeq0.65$) 
hosts fainter RGB and red clump stars; 
and the right one (at $(J-K)\simeq0.9$) corresponds to 
low-mass dwarfs with $M\la0.6$~$M_\odot$. The vertical 
development of these strips in the CMD 
(Fig.~\ref{fig_comp0}a) simply reflects the large range 
of distances involved. 

We notice that disk stars younger than $\sim4$~Gyr 
($M\ga1.3$~$M_\odot$) do not populate significantly this diagram, 
since they are found preferentially at small scale heights 
and hence at lower galactic latitudes and 
brighter apparent magnitudes (i.e.\ $K\la10$).
The contribution of halo stars (black points) 
is also of minor importance, accounting for just a handful of points 
in the synthetic CMD, mostly at fainter apparent magnitudes 
(i.e.\ $K\ga13$), due to their large mean distances.

Finally, an important point is that field stars 
are not expected to reach colours redder
than $(J-K)\simeq1.0$, so that they hardly  contaminate
the CMD features produced by the LMC population of AGB stars.
Their intepretation is our next goal.

\subsubsection{The LMC population}

The LMC population has been initially described by means of a very 
simplified model, under the assumptions of
i) a fixed distance of 52.3 Kpc (Girardi \& Salaris
2001), ii) constant SFR {from  15 Gyr ago to now}, and iii) a 
constant metallicity, $Z=0.008$, at any age. The
latter choice is made  for the sake of simplicity, since
it requires the use of a single set of evolutionary tracks, 
hence {relieving} us, for the moment, from the effort to compute complete 
sets of TP-AGB models for several metallicities\footnote{Extended sets of 
TP-AGB tracks and isochrones for a wide range of metallicities are
being computed and are soon to become available.}. Anyway, the
quality of our results should not depend much on these 
assumptions\footnote{In one of our test cases, 
we used TP-AGB tracks interpolated within an extended
grid of 5 different metallicities, together with a more 
realistic AMR (from Pagel \& Tautvaisiene 1998). No
significant difference was noticed in the resulting CMDs.}.

In order to get a number of LMC giants comparable to the 
observed {sample} (Fig.~\ref{fig_cmdfinal}), we have to
simulate a total star formation of $5\times10^8$~$M_\odot$ 
during the entire LMC history. Of course, this is just a
rough estimate of the mass consumed to form stars, 
its value depending on the adopted assumptions 
about SFR (constant), IMF (from Kroupa 2001), 
and size of the simulated region  
(i.e. within the innermost $4^\circ$ of the LMC field).
Anyway, it is interesting to notice that
this number is just an order of magnitude lower 
than the total LMC mass (including dark matter; see  
van der Marel et al. 2002), and comparable to the present
LMC H\,{\sc i} mass (Staveley-Smith et al. 2003).

The simulated LMC population draws
a few almost vertical features in the CMD (Fig.~\ref{fig_comp0}b), 
mainly populated by: young and 
intermediate-age main-sequence stars (with $(J-K)<0.5$), 
intermediate-mass stars in the stage of
core-He burning (the long diagonal sequence at $(J-K)\simeq0.8$),
intermediate-age and old RGB stars of larger luminosities  
(the prominent sequence at $(J-K)\simeq1$ and with $K>12$), 
and  intermediate-mass stars  during the early-AGB phase
(the weaker sequence departing slightly from the left of 
the RGB-tip, with the same inclination as the RGB). 
All these features (marked with blue dots) 
correspond to O-rich stars in evolutionary stages
previous to the TP-AGB.

The O-rich TP-AGB stars (marked with cyan dots)
trace two well-defined sequences: 
the first consists of intermediate-mass stars that somewhat extend
the plume of early-AGB stars towards higher luminosities;
the second defines a clear strip located directly above the RGB-tip,
corresponding to low-mass stars. 
Finally, The C-rich TP-AGB stars 
(marked with red dots) distribute above and to the right of these latter 
sequences. 

Combining both panels of Fig.~\ref{fig_comp0}, it turns out that 
our initial simulations explain quite well 
the several ``fingers'' displayed by the 2MASS
data of Fig.~\ref{fig_cmdfinal}, but for one major discrepancy:
the location of C-stars, predicted by these particular models,
simply align above the sequence of LMC O-rich giants, failing to 
form the red tail. As for the other CMD features, the 
agreement with data is indeed quite good, and our 
interpretations substantially agree with those advanced by 
Nikolaev \& Weinberg (2000).

\section{Probing the red tail of C-stars}
\label{sec_probing}

In the following we will focus on the region of our simulated 
diagrams populated by TP-AGB stars, with the aim {of casting
 light on} the main cause that produces the
observed red tail of C-stars in near-infrared colors.
We investigate two possible factors separately, namely: the 
colour--$T_{\rm eff}$ relations (derived from empirical 
calibrations), and the evolution in
effective temperatures (derived from stellar models).

\subsection{Testing different colour--$T_{\rm eff}$ relations}
\label{sec_testcolours}

\begin{figure*}
\begin{minipage}{0.48\textwidth}
\includegraphics[width=\textwidth]{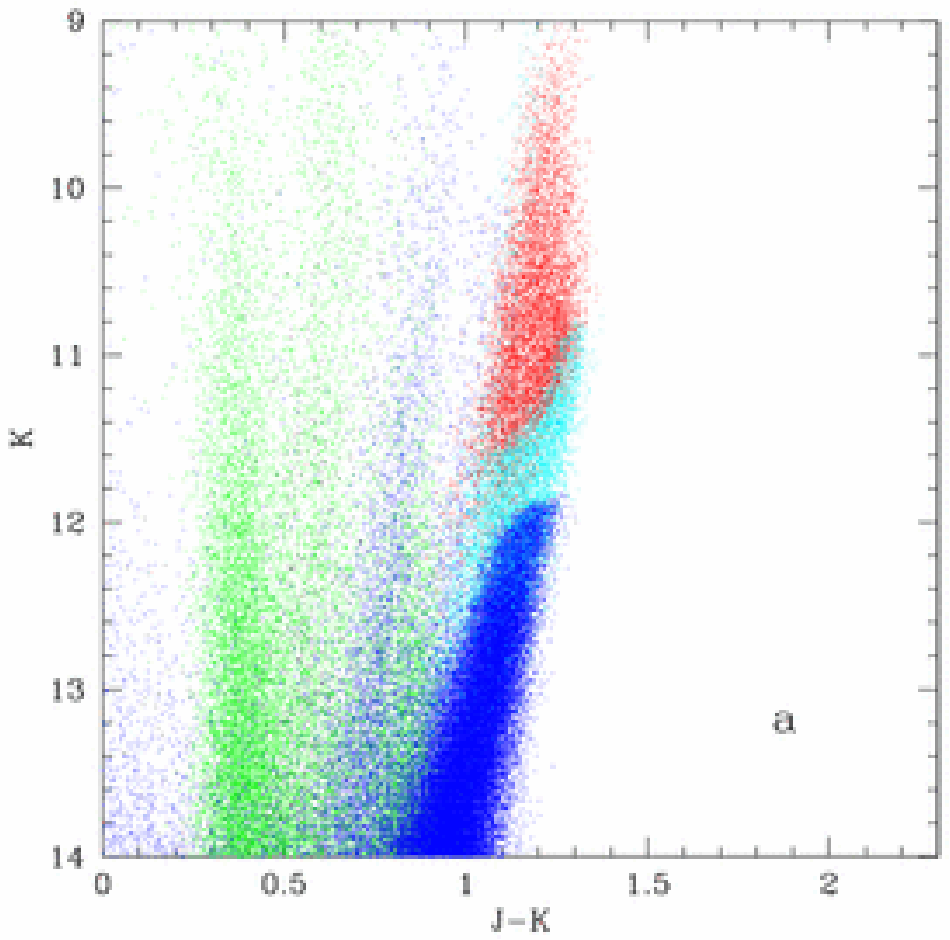}
\end{minipage}
\hfill
\begin{minipage}{0.48\textwidth}
\includegraphics[width=\textwidth]{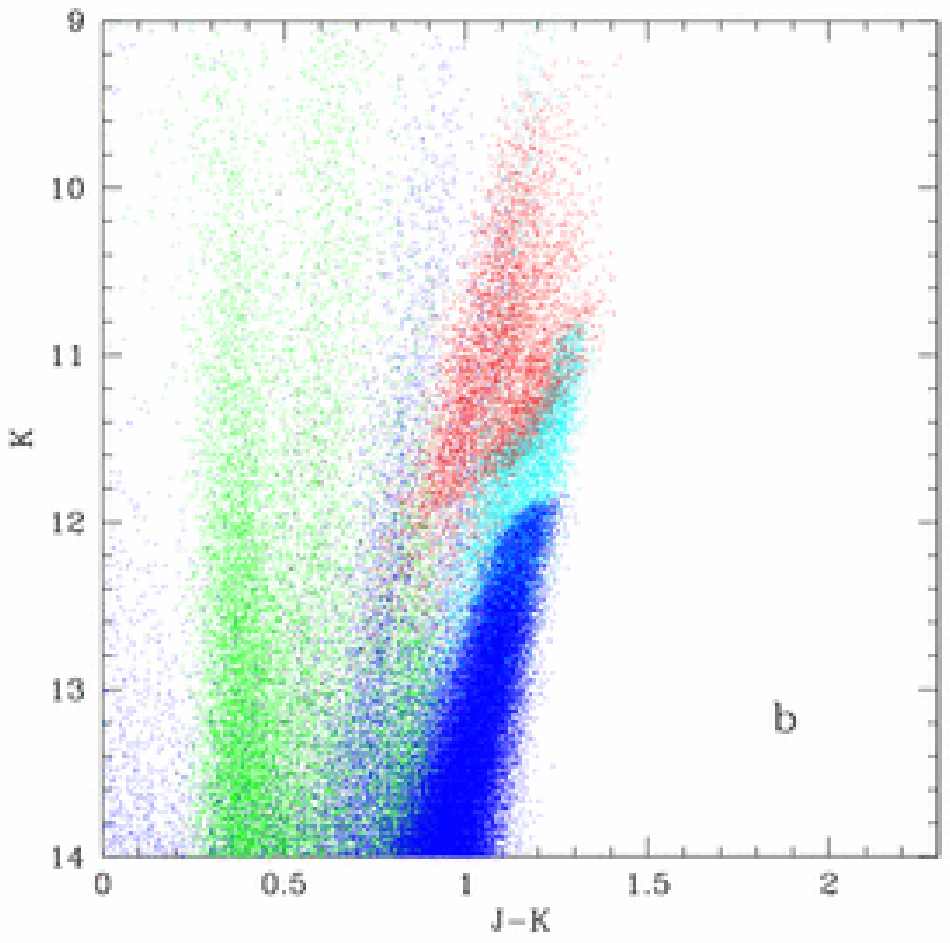}
\end{minipage}
\\
\begin{minipage}{0.48\textwidth}
\includegraphics[width=\textwidth]{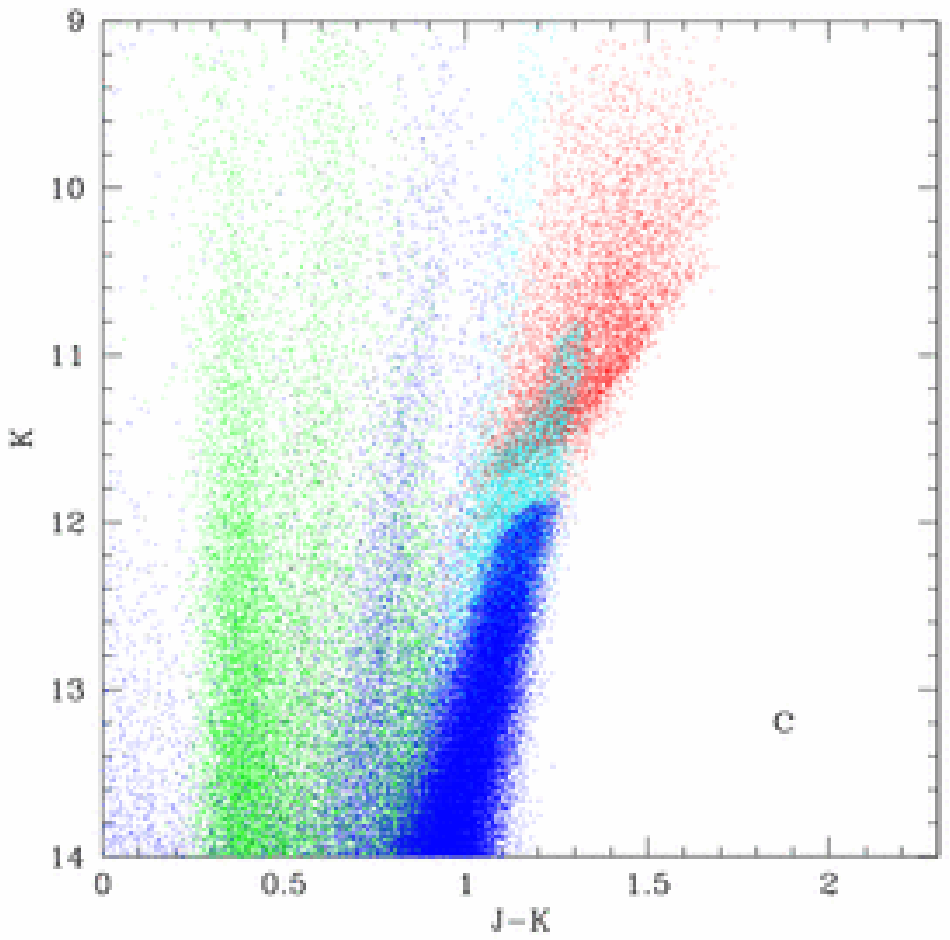}
\end{minipage}
\hfill
\begin{minipage}{0.48\textwidth}
\includegraphics[width=\textwidth]{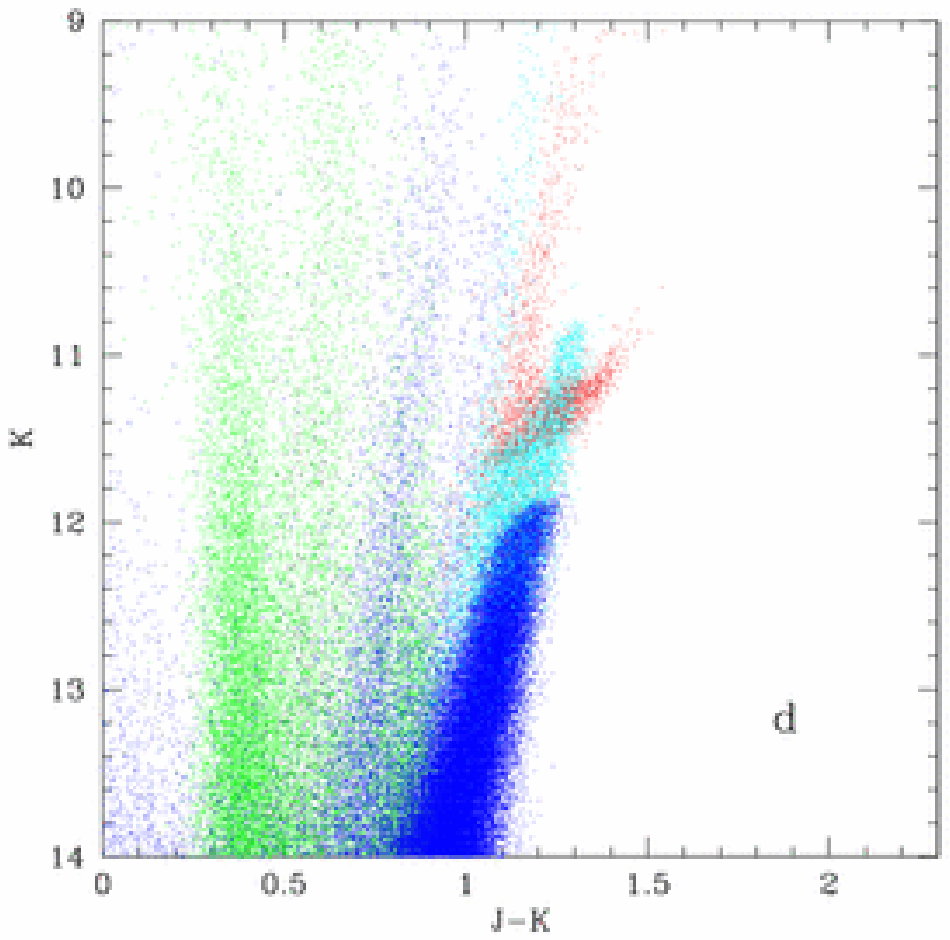}
\end{minipage}
\caption{ 
Simulated CMDs based on TP-AGB tracks with 
$\kappa_{\rm fix}$ prescription for molecular opacities. 
The simulations includes both 
the Galaxy foreground and the LMC population. The coloured dots mark 
several kinds of stars: foreground stars in the Galaxy disk (green)
and halo (black); LMC giants before the TP-AGB phase (blue), 
and separated into O-rich (cyan) and C-rich stars
(red) during the TP-AGB phase. Note that the C-star 
population distribute almost vertically over the red giant branch. 
Different $(J-K)$ vs. $T_{\rm eff}$ transformations are tested in 
the simulations of the C stars:
{\bf Panel a:} The same as for M stars (based on Fluks et al. 1994);
{\bf Panel b:} Bergeat et al. (2001) one;
{\bf Panel c:} A fitting relation involving C/O 
(Eq.~\protect\ref{eq_jkc}). 
A modest inclination appears in the C-star sequence. 
{\bf Panel d:} The same as in Panel C, but plotting only 
the data with a maximum C/O value of 2.0. }
\label{fig_comp1}
\end{figure*}

Let us start by analysing  the effect on our simulations 
of different  $(J-K) - T_{\rm eff}$ transformations. 
It is important to remark that all computations 
discussed in this section are carried
out by adopting TP-AGB tracks for solar-scaled $\kappa_{\rm fix}$
opacities. Results are shown in Fig.~\ref{fig_comp1}.

As {a} first step,  in Sect.~\ref{sssec_ctr1}
we consider the sharp dichotomy between 
colour--$T_{\rm eff}$ relations derived from observed
M-type stars (e.g. Fluks et al. 1994) and C-stars 
(e.g. Bergeat et al. 2001).
Such differences essentially reflect the abrupt change 
in the dominant molecular species -- producing the line
blanketing -- when passing from an O-rich spectrum (e.g. TiO, VO, H$_2$O)
to a C-rich one (e.g. C$_2$, CN, SiC)
 
Then, within the class of C-stars, in Sect.~\ref{sssec_ctr2} we
attempt to account, in a simple way, for 
the possible composition dependence (related to the extent of
carbon enrichment) of colour--$T_{\rm eff}$ relations.

\subsubsection{A $T_{\rm eff}$ -- $(J-K)$ relation valid for 
Galactic C-stars}
\label{sssec_ctr1}

Figure \ref{fig_comp1}a shows our initial simulation:
The TP-AGB models are calculated with  $\kappa$-fix molecular opacities,
and the conversion from $T_{\rm eff}$ to $(J-K)$ is based on 
Fluks et al.\ (1994) empirical spectra and temperature scale 
for nearby M giants (see Girardi et al.\ 2002 for details).
This means that a relation valid for O-rich stars has been applied
also to C-stars. The discrepancy with observed data is obvious:
C-stars just align above the sequence of O-rich stars, and do not
develop any hint of red tail.

\begin{figure*}
\begin{minipage}{0.48\textwidth}
\includegraphics[width=\textwidth]{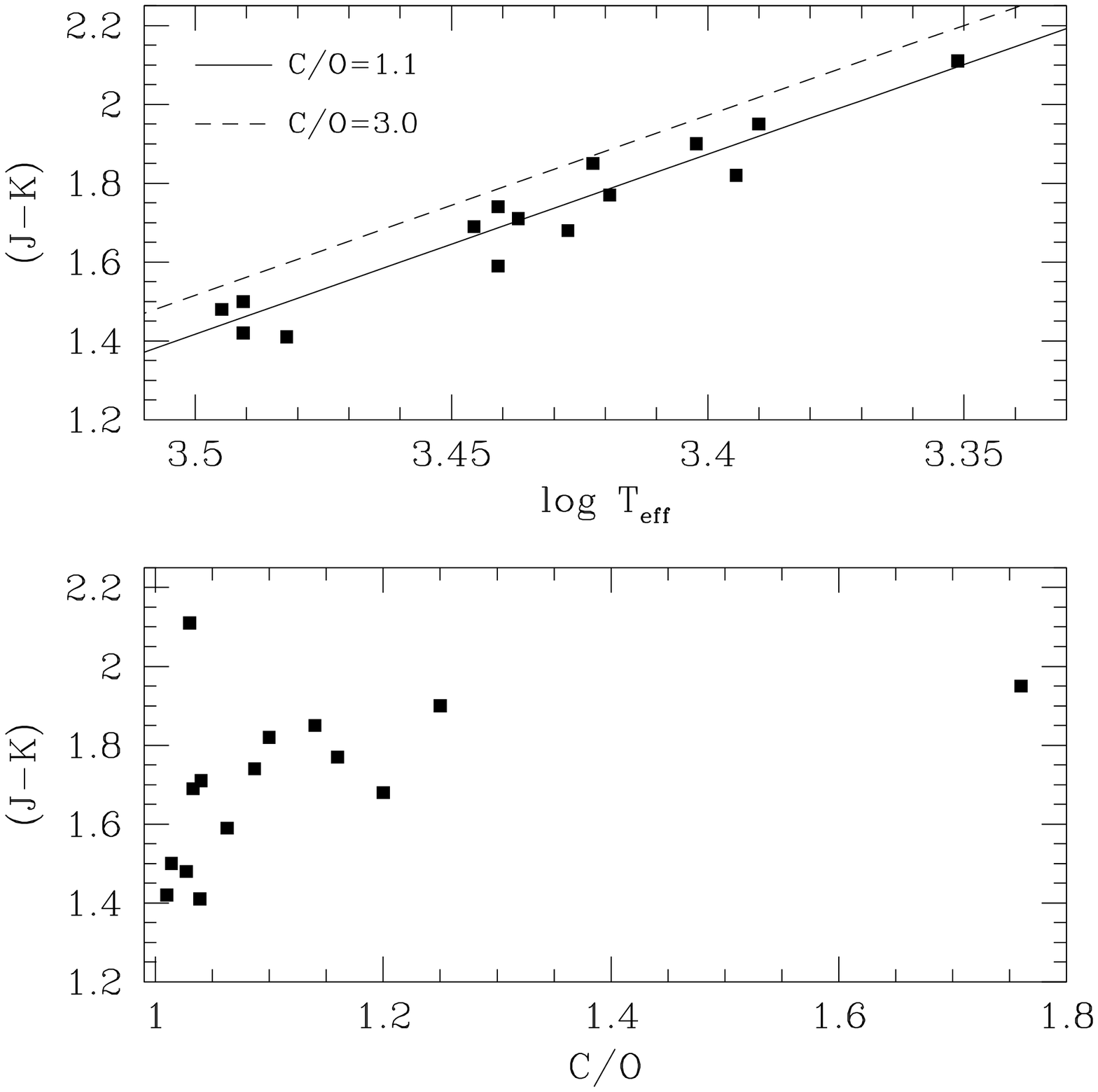}
\end{minipage}
\hfill
\begin{minipage}{0.48\textwidth}
\includegraphics[width=\textwidth]{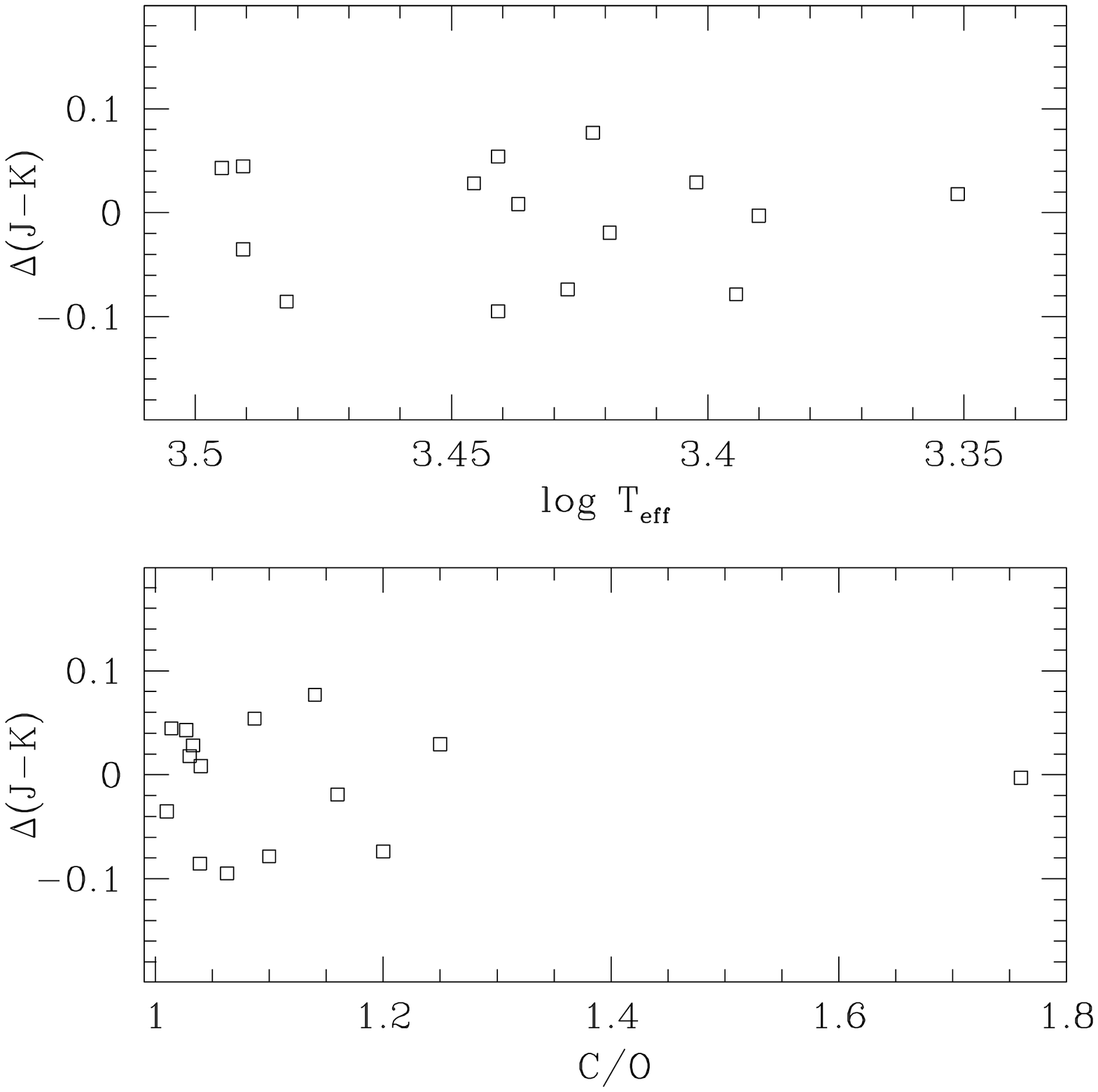}
\end{minipage}
\caption{{\bf Left panels:} $(J-K)$ vs. $T_{\rm eff}$ 
and  $(J-K)$ vs. C/O relations for Galactic C-stars. 
Observed data (filled squares) are from Bergeat et al. 
(2001, objects common to their tables 4 and 13). 
In the upper panel, lines show the predictions 
of the fit relation given by Eq.~(\protect\ref{eq_jkc}),
assuming a constant C/O ratio, equal to 1.1 (solid) and 3.0 (dashed).
{\bf Right panels:} 
The residuals of fitting Eq.~(\protect\ref{eq_jkc}) to the data,
plotted as a function of both $T_{\rm eff}$ and C/O. 
The data is well reproduced by this relation to within $0.1$ 
magnitudes.
}
\label{fig_jktefco}
\end{figure*}

As a first attempt to include proper relations valid for
C-stars, we apply the Bergeat et al.\ (2001) formulas, that 
give $K$-band bolometric corrections and $(J-K)$ as a function of 
$T_{\rm eff}$. These fitting relations are derived from a sample of
Galactic C-stars. {In this context we would like to note that the 
Bergeat et al.\ (2001) relationship is essentially the same as that of
Bessell et al. \ (1981).}  The resulting simulations are shown in
Fig.~\ref{fig_comp1}b. With respect to the previous case,
the sequence of C-stars is shifted to the blue, and not to the
red as required to explain the data.

In this and other plots we show next, we have also adopted the
empirical $K$-band bolometric corrections for C stars 
presented by Frogel et al.\ (1990).
They differ little (by $\sim0.1$ mag) from those 
independently derived by Bergeat et al.\ (2001), and yield 
closely similar simulated CMDs.

\subsubsection{The effect of molecular blanketing for C-stars}
\label{sssec_ctr2}

{Molecular blanketing is responsible for the distinction 
between M- and C-spectral features and  
also affects the infrared  colours of C-stars themselves }
(e.g.\ Cohen et al.\ 1981; Loidl et al.\ 2001).
In this latter case, the extent of molecular blanketing
is essentially related to the amount of carbon available
to form molecules other than CO.

In their study of Galactic and LMC C-stars, Cohen et al.\ (1981) 
pointed out that the observed mean $(J-K)$ colour increases steadily
as the carbon abundance class -- defined by the C$_2$ band strength -- 
increases (from say C,2 to C,5). Moreover, the authors showed  
that, for a given effective temperature, redder $JHK$ colours 
are predicted by model atmospheres with larger C$_2$ and CN enhancements. 

In our work we attempt  
to account, in a simple way, for the effect of molecules on the 
infrared colours of C-stars, by introducing a C/O-dependence into
the empirical $(J-K) - T_{\rm eff}$ relation by Bergeat et al. (2001).
The C/O data for Galactic C-stars are taken from Lambert et al. (1986).
Figure~\ref{fig_jktefco} displays the observed  trends of the $(J-K)$ colour as 
a function of $T_{\rm eff}$ and C/O ratio. 
The linear bi-parametric fit relation to the data is:
\begin{equation}
(J-K) = 17.32 - 4.56 \log T_{\rm eff} + 0.052\,\, {\rm C/O}
\label{eq_jkc}
\end{equation}
We see that the $(J-K) - T_{\rm eff}$ relation 
(top-left panel of Fig.~\ref{fig_jktefco}) is quite narrow for
Galactic C-stars, and it is well reproduced by our fitting formula
for C/O$=1.1$, in agreement with the observational finding that
most Galactic carbon stars have low C/O ratios, not much larger than
one (Lambert et al. 1986). 
Moreover, despite some observed scatter 
and the small number of data points, 
we also note that there is a positive correlation between
the $(J-K)$ colour and the C/O ratio (bottom-left panel of 
Fig.~\ref{fig_jktefco}).
This trend is included in our fitting relation as it predicts  
a systematic shift towards redder $(J-K)$ colours for e.g. C/O$=3.0$, 
compared to the case  C/O$=1.1$ (top left panel).

Overall, Eq.~(\ref{eq_jkc}) fits the data to better than 0.1 mag in
$(J-K)$, as indicated by the residuals in the right panels of
Fig.~\ref{fig_jktefco}.

\begin{figure}
\includegraphics[width=\columnwidth]{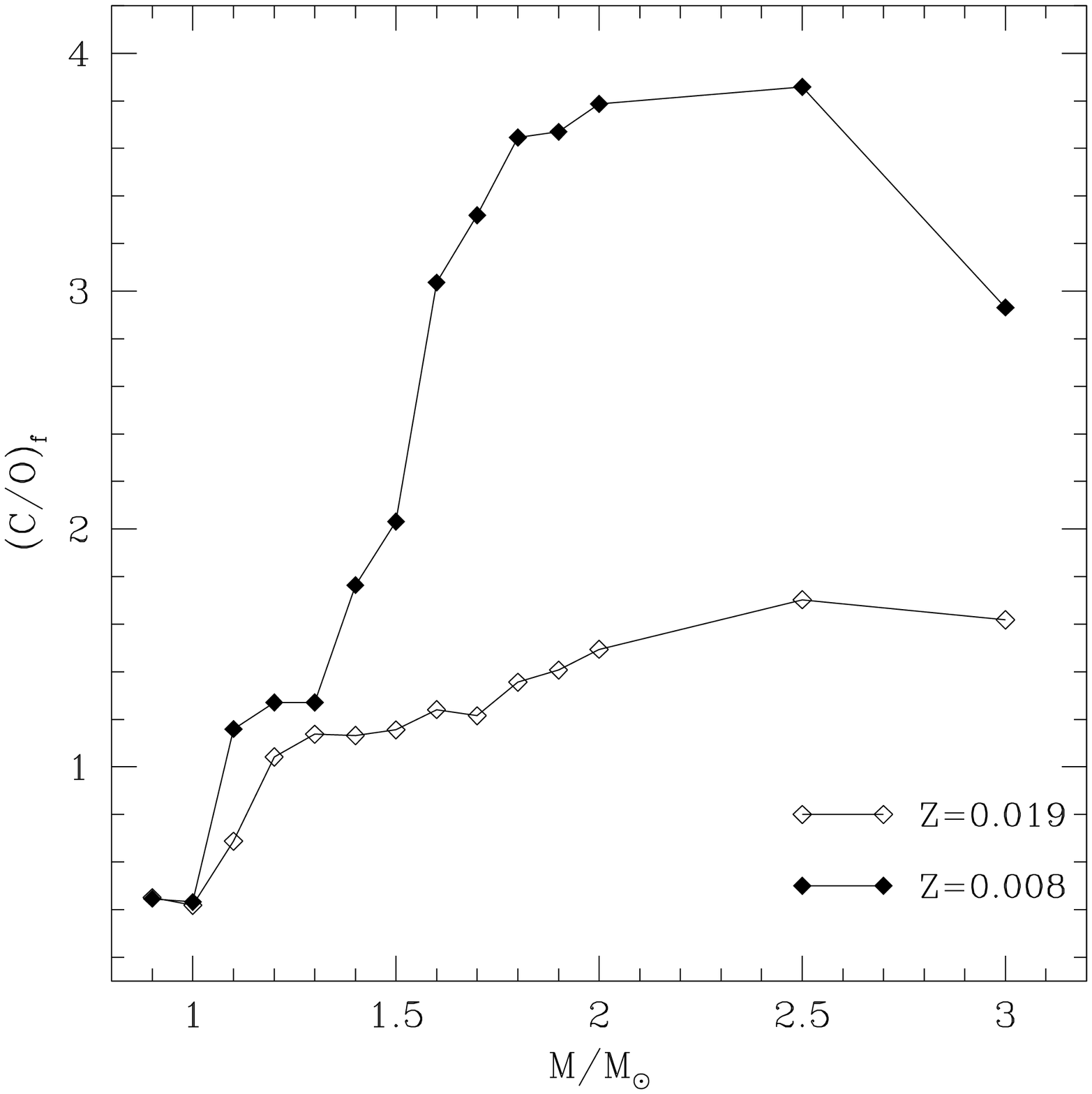}
\caption{Final C/O ratio (at the end of the TP-AGB phase) 
as a function of the stellar mass (at the onset
of the TP-AGB phase), predicted by synthetic models with 
variable molecular opacities $\kappa_{\rm var}$, and 
parameter set ($\lambda=0.50$, $\log T_{\rm b}^{\rm dred} = 6.4$).
Results for solar metallicity ($Z=0.019$) are compared to those for
LMC composition ($Z=0.008$).
}
\label{fig_coz}
\end{figure}

Actually, in a recent study Matsuura et al. (2002) have pointed out that
the molecular features in the spectra of LMC C-stars are consistent 
with C/O ratios larger than those of Galactic C-stars.
This empirical finding completely supports the results of our TP-AGB 
calculations, as shown in Fig.~\ref{fig_coz}. The same figure 
indicates that, in order to model the LMC population, we will have
to apply Eq.~(\protect\ref{eq_jkc}) to C/O values as high as 4, that
is beyond the maximum observed values of Galactic C-stars. 

Thus, by using Eq.~(\ref{eq_jkc}) in our simulations, we get the
synthetic CMD of Fig.~\ref{fig_comp1}c. As it can be noticed,
the dependence of $(J-K)$ on the C/O ratio now causes the sequence
of C-stars to be significantly redder, reaching $(J-K)=1.7$.
However, the C-stars are still largely distributed
among and above the O-rich giants, and no clear red tail is drawn.

{If we limit consideration} to  models with C/O$<2$, i.e. plotting
only those stars for which Eq.~(\ref{eq_jkc}) is strictly valid, 
we get Fig.~\ref{fig_comp1}d. In this case, we can notice that
the low-C/O C-stars contaminate significantly the CMD region 
corresponding to O-rich AGB stars. On the contrary, such a
high fraction of C-stars mixed up with M-stars is not present
in observed samples.
It is clear that this discrepant feature would remain 
for whatever hypothetical $(J-K) - T_{\rm eff} -$ C/O 
that could push somehow the models to reach $(J-K)$ colours as high as $2.0$.

In conclusion,  our test calculations seem to indicate 
that the principal cause of the failure of models
with solar-scaled molecular opacities in 
reproducing the red tail of carbon stars 
should not be ascribed  to the $(J-K)-T_{\rm eff}$ relation.

\subsection{Testing different TP-AGB tracks}
\label{sec_testtracks}

If we assume that Eq.~(\ref{eq_jkc}) reasonably well describes
the $(J-K)$ colours of C-stars, as supported by the low fitting 
residuals (of $\la0.1$~mag, see
the right panels of Fig.~\ref{fig_jktefco}),
we are left with just one alternative:  
redder $(J-K)$ colours could be attained only by 
assigning  lower $T_{\rm eff}$ to C-star models.
Moreover, in order to locate the C-star models onto  
a different branch from that of the O-rich models, 
a net $T_{\rm eff}$ segregation between the two classes 
turns out to be necessary. These requirements are indeed 
fulfilled by Marigo (2002) TP-AGB models 
with variable molecular opacities.

\begin{figure}
\includegraphics[width=\columnwidth]{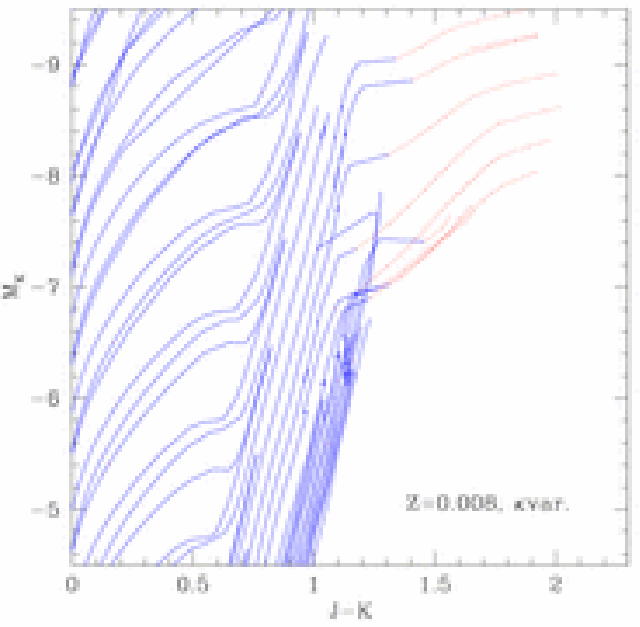}
\caption{$Z=0.008$ theoretical isochrones for TP-AGB models
computed with variable molecular opacities. The dredge-up
parameters are the same as in the models of 
{Fig.~\protect\ref{fig_isockfix}}. The dotted lines 
mark the isochrone sections corresponding to C-stars. 
The conversion to the $(J-K)$ photometry is now based on 
Eq.~(\protect\ref{eq_jkc}).
}
\label{fig_isockvar}
\end{figure}

Figure~\ref{fig_isockvar} illustrates a set of 
$Z=0.008$ isochrones based on  
Marigo's  (2002) models. Similar isochrone sets 
have been computed for a large variety of dredge-up 
parameters and metallicities, and will be extensively 
described in a future paper. The conversion to $(J-K)$ colours 
for C-star models is now based on Eq.~(\ref{eq_jkc}). 
It is also interesting to notice that in this case  
we could not even have applied 
the same colour transformations derived for M-stars, 
since the effective temperatures of C-star models with $\kappa_{\rm
var}$ are already much cooler than the validity lower-limits of such relations.

\subsubsection{The sole effect of variable molecular opacities}
\label{sssec_konly}

\begin{figure*}
\begin{minipage}{0.48\textwidth}
\includegraphics[width=\textwidth]{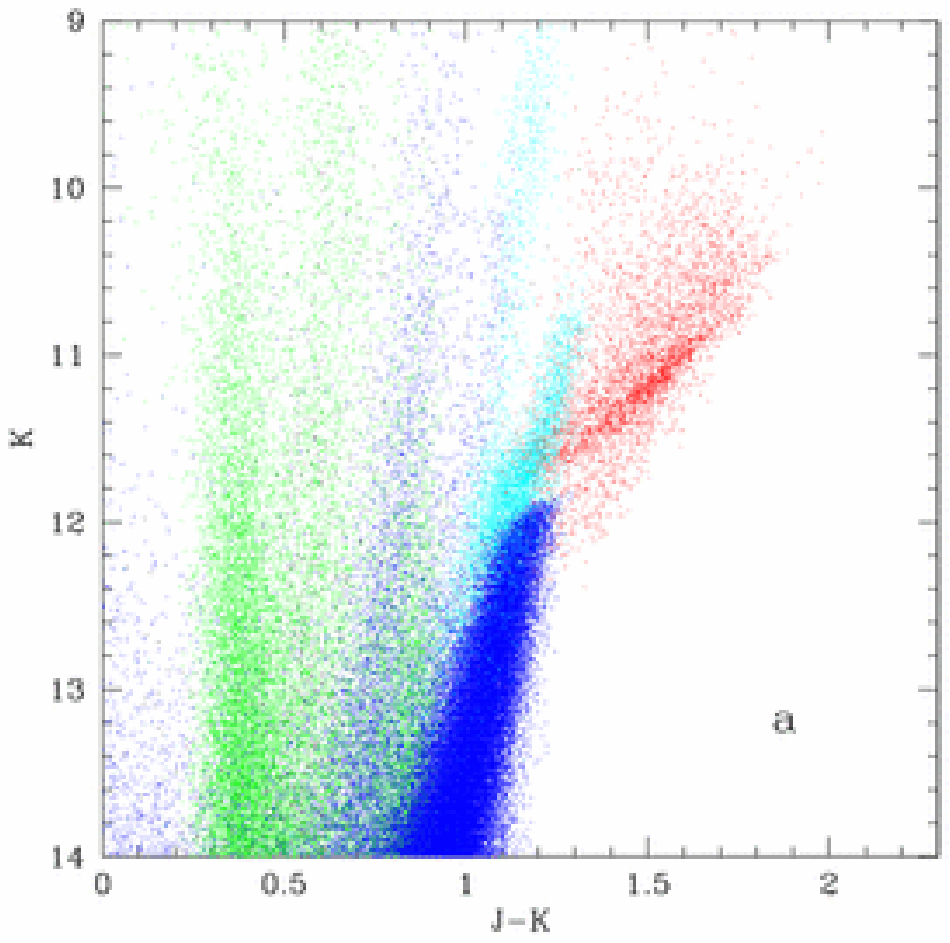}
\end{minipage}
\hfill
\begin{minipage}{0.48\textwidth}
\includegraphics[width=\textwidth]{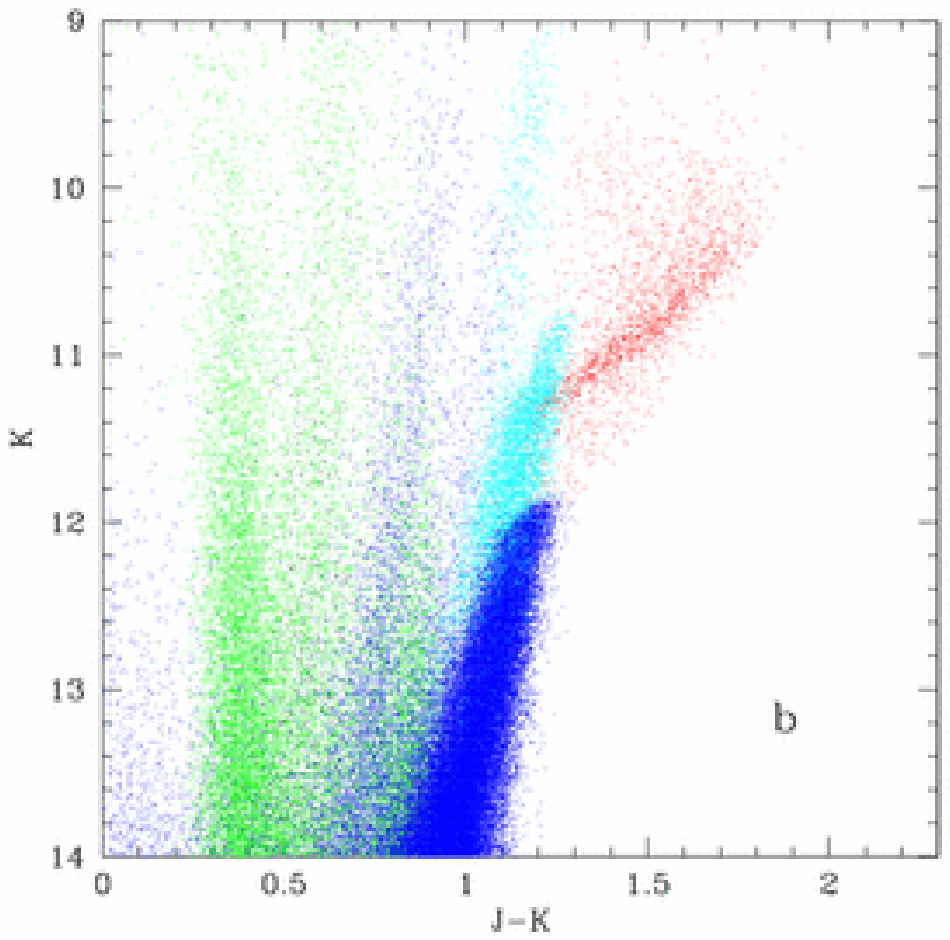}
\end{minipage} \\
\begin{minipage}{0.48\textwidth}
\includegraphics[width=\textwidth]{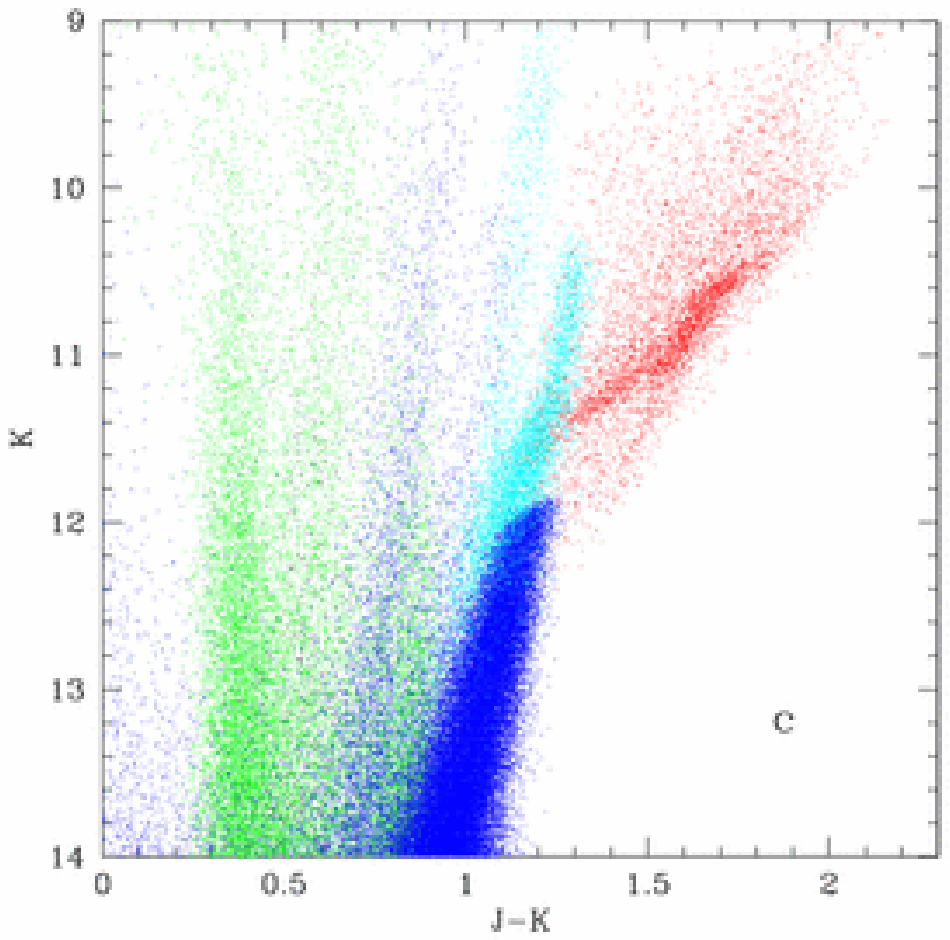}
\end{minipage}
\hfill
\begin{minipage}{0.48\textwidth}
\includegraphics[width=\textwidth]{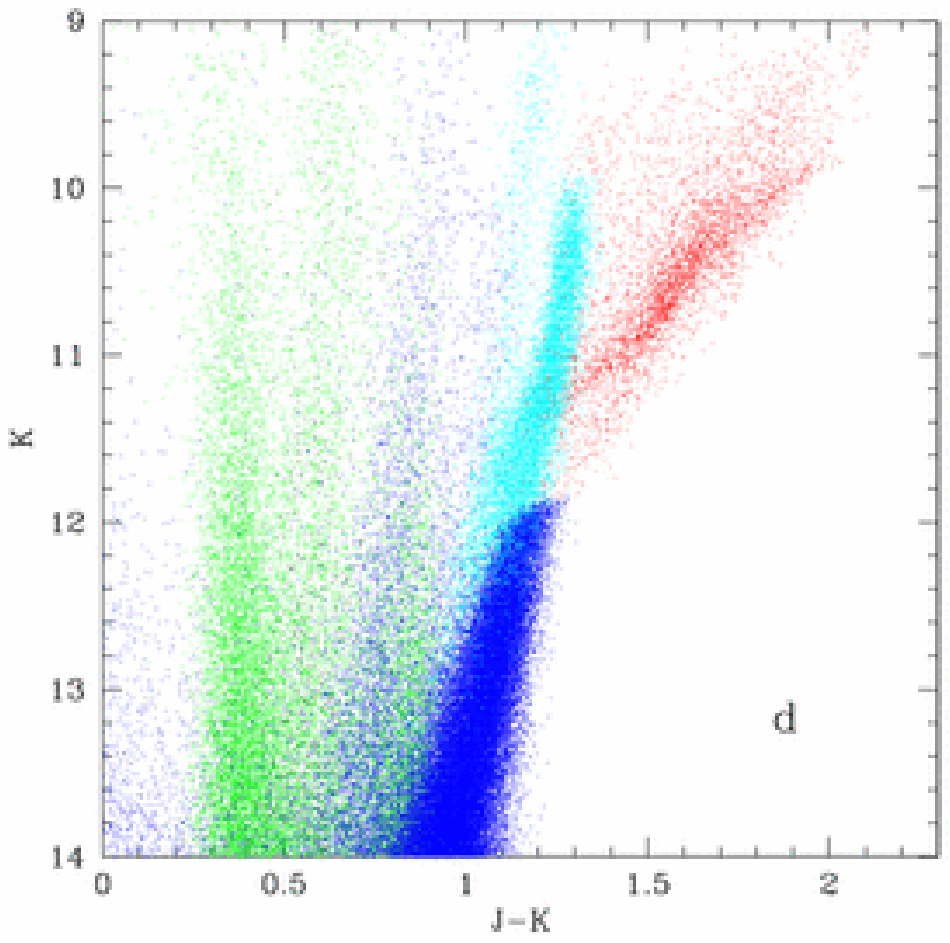}
\end{minipage}
\caption{Simulated H-R diagram based on TP-AGB tracks with variable
molecular opacities (i.e. Marigo 2002). Note the departure of the C-star 
population away from the red giant branch  towards redder colours.  
The mass loss prescriptions is taken from Vassiliadis \& Wood (1993), 
assuming that AGB stars are either fundamental-mode or first
overtone pulsators. 
{\bf Panel a:}  case $\log T_{b}^{\rm dred}=6.4$, fundamental mode. 
{\bf Panel b:}  case $\log T_{b}^{\rm dred}=6.45$, fundamental mode. 
{\bf Panel c:}  case $\log T_{b}^{\rm dred}=6.4$, first overtone. 
{\bf Panel d:}  case $\log T_{b}^{\rm dred}=6.45$, first overtone. 
}
\label{fig_comp2}
\end{figure*}

The effect of TP-AGB models with variable molecular
opacities is illustrated in the simulations of
Fig.~\ref{fig_comp2}. TP-AGB models are computed 
with different assumptions (see also Sect.~\ref{sec_Cstars}) 
regarding the onset of the third dredge-up 
-- controlled by the parameter $T_{\rm b}^{\rm dred}$ --, and the onset 
of the super-wind regime -- depending on the pulsation mode, either
fundamental ($P=P_0$), or first overtone ($P=P_1$). In all cases, the
sequence of C-stars clearly departs from the branch
of O-rich stars. 

In these $\kappa_{\rm var}$ models the excursion towards redder $(J-K)$ 
colours  is mostly caused by the displacement towards cooler  
$T_{\rm eff}$. Instead, the increase of the C/O ratio 
during the TP-AGB evolution should play a smaller role. 
In fact, considering that 
the maximum C/O ratio reached by these models is 
of about 4 (Fig.~\ref{fig_coz}), the C/O-dependence in the
colour transformation of Eq.~(\ref{eq_jkc}) accounts
for at most $0.2$~mag of the total colour excursion
at the end of the C-star evolution, the mean
value being of about 0.1~mag.

In Fig.~\ref{fig_comp2}a, we adopt the same dredge-up parameters 
as calibrated in Marigo et al. (1999), i.e.\ 
$\log T_{\rm b}^{\rm dred}=6.4$ and $\lambda=0.5$. The simulated 
C-star red tail resembles very much the observed one, but for two 
aspects. 

First, the red tail shows up at a too low luminosity.
It starts at $K=11.6$ when $(J-K)=1.2$, whereas the observed tail 
is at least 0.6~mag brighter. This problem can be easily solved 
with a re-calibration of the dredge-up parameters. 
To give an example, Fig.~\ref{fig_comp2}b shows a 
simulation that  makes use of TP-AGB models 
recomputed with $\log T_{\rm b}^{\rm dred}=6.45$. 
This parameter choice has the effect to delay the
onset  of the third dredge-up to higher stellar core masses and hence
higher luminosities. As we can see, now the red tail starts at $K=11.0$ 
when $(J-K)=1.2$, which is about the right luminosity level to fit
the data.

It is worth remarking that in both simulations presented in the
upper panels (a and b) of Fig.~\ref{fig_comp2},
we have assumed that all TP-AGB stars are fundamental-mode pulsators,
meaning that the Vassiliadis \& Wood's (1993) mass-loss formalism 
is applied with $\dot M(P) = \dot M(P_0)$.
This assumption introduces the second problem of the models, 
which is discussed below.

\subsubsection{Extending the red tail with first-overtone 
pulsators}
\label{sec_furtherextending}

The mentioned second theoretical difficulty resides 
in the extension of the red tail, 
which in both simulations (Figs.~\ref{fig_comp2}a and b) 
clearly reaches $(J-K)=1.8$, but hardly $(J-K)=2.0$ as in the observations.
{Although this point could be considered as a minor discrepancy 
(at least compared with the $\kappa_{\rm fix}$ cases illustrated in 
Fig.~\ref{fig_comp1}), it deserves some attention. There are at 
least two possible ways out.  The first is that we have not considered 
the effect of circumstellar reddening on the colours of C-stars. 
This should be ascribed to thick circumstellar material ejected 
by the coolest C-stars during an extreme mass-loss phase.
In fact, a few LMC C-stars are known to 
extend up to $(J-K)\sim6$ (Cioni et al.\ 1999; 
Nikolaev \& Weinberg 2000); they are among the so-called ``obscured'' 
or ``dust-enshrouded'' C-stars identified with IRAS sources
(see van Loon 1999), which are found to undergo heavy mass-loss. 
The question, however, is if this sort of
self-reddening is already effective in red tail C-stars 
with say $(J-K)_0\la1.8$, being able to shift their colours 
up to $(J-K)=2.0$.
This point has to be investigated with the aid of more detailed 
modelling.

The second explanation stands on the assumption that
the dominant pulsation mode of AGB stars is the
first overtone (with period $P=P_1$).
This would allow us to get simulated colours further 
increased by 0.2~mag, while keeping the same $T_{\rm eff}$ vs. $(J-K)$
relation.} 
This is illustrated in the bottom panels of
Fig.~\ref{fig_comp2}.  For these simulations we use TP-AGB models
computed with the same dredge-up parameters as in the 
upper panels, but assuming that Mira and SR variables 
are first overtone pulsators. 
Under this hypothesis, variable 
AGB stars should have shorter pulsation
periods for given stellar $L$ and $T_{\rm eff}$, compared to the 
fundamental-mode counterparts.

As a consequence, {by} virtue of the  empirical  
positive correlation between the mass-loss rates and  the periods
of Mira stars, we expect that for the shorter first
overtone periods ($P_0/P_1 \sim 2.2$) the  onset of the super-wind regime
is postponed, thus implying that the end of the AGB evolution 
is delayed. We quantify this effect by adopting the Vassiliadis 
\& Wood's (1993) formalism with $\dot M(P)=\dot M(P_1)$
(Figs.~\ref{fig_comp2}c,d). 
In this way we find that C-star models
are allowed to reach slightly higher $L$ and 
lower $T_{\rm eff}$, and their lifetimes increase accordingly. 
{This latter effect is also relevant if we consider
that longer lifetimes are needed to
better reproduce the observed C-star counts in LMC clusters 
(Girardi \& Marigo 2003, and work in preparation).}

In Fig.~\ref{fig_comp2}c, we can notice that now the red tail 
actually extends up to $(J-K)=2.0$. However, it still 
appears at too faint luminosities,
due to the low value assumed for $T_{\rm b}^{\rm dred}$ 
(see remarks in Sect.~\ref{sssec_konly}).
Instead, for a larger $T_{\rm b}^{\rm dred}$, 
the consequent later onset of the
third dredge-up is able to shift the entire red tail to the
right luminosities, as shown  in Fig.~\ref{fig_comp2}d.

From a comparison with  our reference plot of
Fig.~\ref{fig_cmdfinal}, 
we can conclude that the simulation of Fig.~\ref{fig_comp2}d 
best reproduces the observed morphology of the red tail. 
On the other hand,  the introduction of
the first overtone hypothesis produces an undesirable
effect in the sequence of old O-rich stars: they now extend
to much higher luminosities (up to $K=10.4$ and $K=10.0$ in
Figs.~\ref{fig_comp2}c and d, respectively) than observed
(up to $K=11.2$ in Fig.~\ref{fig_cmdfinal}).

As a matter of fact, nowadays the observational scenario 
of long-period variables (LPVs) appears quite complex, mainly due to the
observed mixture of different pulsation modes, and
the lively controversy about their assignment (see e.g. Wood et al. 1999, 
Feast 1999, Whitelock \& Feast 2000, Cioni et al. 2001, Noda et
al. 2002).
The situation gets even more {intricate} if one considers 
the evidence that some LPVs are found to be multi-period pulsators 
(Bergeat et al. 2002), and 
the possibility of an evolutionary path
across different period-luminosity relations, so that LPVs
may switch pulsation modes while evolving along the AGB
(Cioni et al. 2001).

Without entering the open debate on the dominant pulsation mode, we
perform a very simple test on the basis of purely theoretical
{considerations}.  Considering that the red tail of C-stars is somewhat
better reproduced with the first-overtone assumption which, instead,
does not seem to suitably describe the O-rich branch, we assume a
mixed population of fundamental-mode and first-overtone pulsators,
with the following composition:
\begin{enumerate}
\item All stars with mass lower than 1.3~$M_\odot$, and 
larger than 3.5~$M_\odot$, are fundamental-mode pulsators
just because they do not evolve into a C-star phase;
\label{item_1}
\item The remaining stars are assigned {\em either} as 
fundamental-mode {\em or} as first-overtone pulsators.
We assume a mix of 50 percent between the two types.
\label{item_2}
\end{enumerate}
%
\begin{figure*}
\begin{minipage}{0.48\textwidth}
\includegraphics[width=\textwidth]{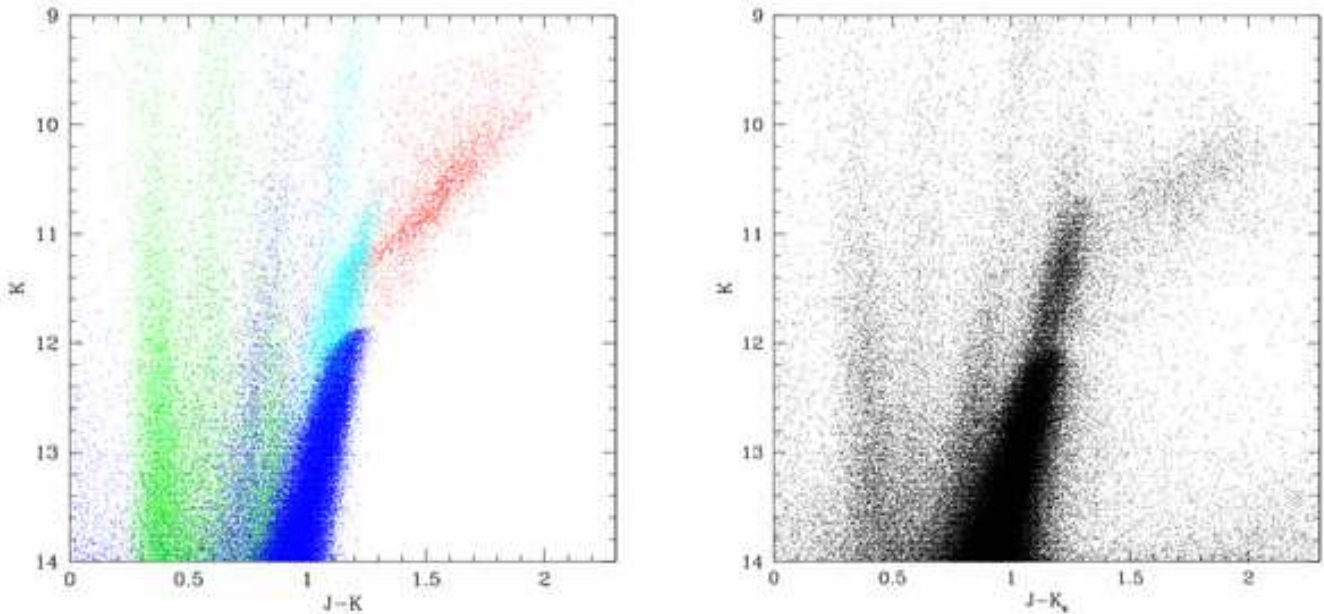}
\end{minipage}
\hfill
\begin{minipage}{0.48\textwidth}
\includegraphics[width=\textwidth]{./kjk_2mass_4deg.eps}
\end{minipage}
\caption{{\bf Left panel:} 
Simulated CMD based on TP-AGB tracks with variable
molecular opacities (i.e. Marigo 2002) and 
$\log T_{\rm b}^{\rm dred}=6.45$,
assuming an half-and-half mixture of fundamental-mode and 
first-overtone pulsators for the stars which evolve into the
C-star phase. 
{\bf Right panel:} The same diagram for the selected 2MASS data.
Notice the very good description of all CMD features,
including the C-star red tail (in red) and the plume 
of O-rich low-mass TP-AGB stars (in cyan).}
\label{fig_comp3}
\end{figure*}
%
Assumption (\ref{item_2}) may be not realistic 
enough, because the simulated stars do not switch between
pulsation modes as they evolve, but they can just follow one
between two different evolutionary paths.  
If this is an oversimplification of the problem --
unavoidable because the mechanisms leading to change the 
pulsation mode cannot be presently modelled  --
at the same time it is a resonable and instructive experiment.
In Fig.~\ref{fig_comp3} we present the resulting 
simulation, compared with the selected 2MASS data. 
We can notice, in fact, that the {resulting sequence of O-rich giants
is noticeably shortened in luminosity, as expected}. 

We may therefore conclude that now our final simulation
actually meets 2MASS and DENIS observations, as 
{\em all CMD features are well reproduced by the models,
including the C-star red tail and the plume of
O-rich low-mass TP-AGB stars}.

\section{Concluding remarks}
\label{sec_conclusion}

In this paper, we have shown that 
most features seen in the 2MASS and DENIS $K_{\rm s}$ vs. 
$(J-K_{\rm s})$ diagram are quite well described by
our present stellar models.
In particular, the observed tail
towards the reddest $(J-K_{\rm s})$ colours
provides a direct and stringent probe to the evolutionary theory of
TP-AGB stars. In this regard, we have pointed out that:
\begin{enumerate}
\item The red tail of carbon stars is missed by 
models in which the TP-AGB evolution is followed
using solar-scaled molecular opacities. 
\item This failure can hardly be caused
by some hypothetical inadequacy in the present-day
$T_{\rm eff}$ vs. $(J-K)$ relations for C-stars. 
Our tests  indicate that in order to reach the observed $(J-K)$
colours of C-stars with ``solar-scaled opacity'' models, 
quite high mean values of the C/O ratio should be
exhibited by {\sl the  bulk of} carbon stars, which
seems unrealistical.
\item Instead, the red tail comes out naturally of 
TP-AGB models which are calculated with variable molecular
opacities (cf. Marigo 2002), i.e. opacities
consistently computed for the current chemical composition of
the stellar envelope.
This can be understood considering that, as the third dredge-up makes
the C/O ratio increase above 1, drastic changes in the 
main sources of molecular opacities cause both  
the transition from M- to C-spectral types, and a
significant decrease of the effective temperature. 
The latter aspect has been so far largely ignored in AGB evolution
models.
Contrary to what claimed by Mouhcine \& Lan\c con (2002), 
the present  study demonstrates that in order to model {\sl properly}
the properties of AGB, in particular C-rich,  
stellar populations, 
one cannot {ignore} 
the large opacity effects have on the effective temperature 
of C-stars, and more generally on the AGB evolution. 
\item Our simulations  with ``variable opacity'' models 
provide a few additional useful indications about the 
TP-AGB evolution. For instance,
models with $\log T_{\rm b}^{\rm dred}=6.45$
seem to reproduce well the $K$-band luminosity 
of red tail C-stars. This means that the onset of 
the third dredge up should occur at slightly larger core masses, 
as compared to previous models by Marigo et al. (1999).
\item Moreover, in the context of the present models,
based on the use of Vassiliadis \& Wood (1993) prescription for
mass loss and an empirical $T_{\rm eff}$ vs. $(J-K)$
relation, we may derive two indications about the pulsation modes of 
variable AGB stars. First, the colour extension of the red tail 
{may imply a significant fraction of 
first-overtone pulsators among C-type stars. However, 
the same extension could be reproduced with fundamental-mode
pulsators if circumstellar reddening starts to be important 
for C-stars with $(J-K)_0\sim1.8$, 
a point that has still to be properly investigated.}
Second, the luminosity extension of the sequence populated by 
O-rich low-mass TP-AGB stars would suggest that most of M-type 
variables are fundamental-mode pulsators.
\end{enumerate}

All these indications are being used to calibrate
our TP-AGB models, from which extended sets of
theoretical isochrones at varying metallicity will be derived and
soon made available. Of course, new models will
be computed for the variable-opacity case only.

\begin{acknowledgements}
We would like to thank our referee Dr.\ M.\ Bessell, and P.R.\ Wood,
for useful remarks that improved the final version of the paper.
We thank M.\ Groenewegen for his fine calibration of the 
parameters used in the Galaxy model and for useful comments.
This publication makes use of data products from the 
Two Micron All Sky Survey, which is a joint project of the 
University of Massachusetts and the Infrared Processing and 
Analysis Center/California Institute of Technology, funded by 
the National Aeronautics and Space Administration and the 
National Science Foundation. The DENIS data is the result of
an extended international collaboration with the use of ESO
telescopes.
We acknowledge financial support from the MIUR-COFIN 2001 project
``The tridimensional structure of the Galaxy'' (number
2001028112).

\end{acknowledgements}

\end{document}